\documentclass[a4paper]{article}
\pdfoutput=1
\usepackage{amsmath}
\usepackage{amsthm}
\usepackage{amssymb}
\usepackage[utf8]{inputenc}
\usepackage{hyperref}
\usepackage{cleveref}
\usepackage{autonum}
\hypersetup{
	unicode,
}
\usepackage{graphicx, subfigure, color}
\graphicspath{{fig/}}
\usepackage{algorithm, algpseudocode} 
\usepackage{mathrsfs} 
\usepackage{authblk}

\newcommand{\x}{\mathbf{x}}
\usepackage{comment}
\usepackage[switch, modulo]{lineno}
\title{A kinetic approach to consensus-based segmentation of biomedical images}

\author[1,2,3]{Raffaella Fiamma Cabini}
\author[4,5]{Anna Pichiecchio}
\author[6,2]{Alessandro Lascialfari}
\author[7]{Silvia Figini}
\author[1]{Mattia Zanella}

\affil[1]{Department of Mathematics, University of Pavia, Via Ferrata 5, 27100 Pavia, Italy}
\affil[2]{INFN, Istituto Nazionale di Fisica Nucleare, Pavia Unit, Via Bassi 6, 27100, Pavia, Italy}
\affil[3]{Euler Institute, USI, Via la Santa 1, 6962, Lugano, Switzerland}
\affil[4]{Department of Brain and Behavioural Sciences, University of Pavia, Via Mondino 2, 27100 Pavia, Italy}
\affil[5]{Advanced Imaging and Radiomic Center, IRCCS Mondino Foundation, Via Mondino 2, 27100 Pavia, Italy}
\affil[6]{Department of Physics, University of Pavia, Via Bassi 6, 27100, Pavia, Italy}
\affil[7]{Department of Social and Political Science, University of Pavia, Corso Carlo Alberto 3, 27100 Pavia, Italy}

\begin{document}
	\maketitle
	%%==================================%%
    %% ABSTRACT %%
    %%==================================%%
	\begin{abstract}
	In this work, we apply a kinetic version of a bounded confidence consensus model to    biomedical segmentation problems. In the presented approach, time-dependent information on the  microscopic state of each particle/pixel includes its space position and a feature representing a static characteristic of the system, i.e. the gray level of each pixel. 
    From the introduced microscopic model we derive a kinetic formulation of the model. The large time behavior of the system is then computed with the aid of a surrogate Fokker-Planck approach that can be obtained in the quasi-invariant scaling.  We exploit the computational efficiency of direct simulation Monte Carlo methods for the obtained Boltzmann-type description of the problem for parameter identification tasks. Based on a suitable loss function measuring the distance between the ground truth segmentation mask and the evaluated mask, we minimize the introduced segmentation metric for a relevant set of 2D gray-scale images.  Applications to biomedical segmentation concentrate on different imaging research contexts.
		\\
		
	\end{abstract}

	%\tableofcontents
	
	%%==================================%%
    %% INTRODUCTION %%
    %%==================================%%
    
    \section{Introduction}\label{sec1}
    
    In image processing and computer vision, image segmentation is a fundamental process to subdivide images in subsets of pixels that has found application in many research contexts \cite{cheng2001color}. In the field of medical imaging, the identification of image subregions is a powerful tool for tissue recognition to  track pathological changes. Image segmentation can help clinical studies of anatomical structures and in the identification of regions of interest, and to measure tissue volume for clinical purposes.

    The main goal of image segmentation is to divide the image into a set of pixel regions that share similar properties such as closeness, gray level, color, texture, brightness, and contrast~\cite{sharma2010automated}. By converting an image into a group of segments, it is possible to process only the important areas instead of studying the entire image. To this end, various computational strategies and mathematical methods have been developed in the last decades. Among them, Neural Networks (NNs)  are one of the most common strategies used in modern image segmentation problems. These techniques can approximate, starting from a series of examples, the nonlinear function between the inputs and the outputs of interest. It has been observed that well trained NNs can achieve good segmentation accuracy even with complex images~\cite{isensee2021nnu, ronneberger2015u, zhou2018unet, hesamian2019deep}. Anyway, the approximation obtained through NNs may require  extensive supervised training procedure. Indeed,  the performances of NNs essentially depend on the adopted training procedure and on the availability of unbiased data  ~\cite{liu2021review}. 
    
    A different approach to image segmentation is based on clustering techniques such as the k–means method, the c–means method, hierarchical clustering method and genetic algorithms~\cite{jain1999data,HAN2017588}. We refer to clustering process as a dynamics to identifying groups of similar data points according to some observed characteristics. These strategies belong to the category of the unsupervised algorithms since they do not require a training procedure and do not depend on training datasets. It can be observed that image segmentation is a clustering process where the pixels are classified into multiple distinct regions so that pixels within each group are homogeneous with respect to certain features, while pixels in different groups are different from each other~\cite{mittal2021comprehensive, frigui1999robust,YU20101889, pizzagalli2019trainable}. 
        
    In this work, following the approach introduced in the recent work~\cite{herty2020mean} for clustering problems, we adopt a mathematical strategy to image segmentation that is based on consensus dynamics of large systems of agents. In this direction, we adopt a kinetic-type approach by rewriting  generalized Hegselmann-Krause (HK) microscopic consensus  models in terms of a binary scheme. The evolutions of aggregate quantities are then obtained through a Boltzmann-type model whose steady state can be approximated by means of a quasi-invariant approach \cite{pareschi2013interacting}. Indeed, in the mentioned scaling, a reduced complexity Fokker-Planck model corresponds, in the zero-diffusion limit, to the mean-field model defined in \cite{herty2020mean}. Suitable steady state preserving numerical methods can be applied to verify the consistency of the approach. 
    
    Following the approach in \cite{herty2020mean}, each pixel is represented by a particle characterized by a time-dependent position vector and a static feature that describes an intrinsic property of the particle, i.e. the gray level of the pixel. Particles interact with each other until they reach the equilibrium state in which they group into a finite number of clusters. Hence, a segmentation mask is generated by assigning the mean of their gray levels to each cluster of particles and by applying a binary threshold. The two main advantages of this segmentation system compared to other standard clustering techniques such as the k–means method, are that the clustering process also takes into account the gray level of the pixels and not just their reciprocal positions and it is not necessary to select in advance the final number of the clusters. 
    
    In the following, we apply a particle-based clustering method to biomedical image segmentation problems. In particular, we will incorporate in the model a non-constant diffusion term which depends on the gray level of the pixel which consent to quantify aleatoric uncertainties in the segmentation pipeline. The aleatoric uncertainties derive from image acquisition processes and can affect the quality of the considered image.

    The Boltzmann-type formulation of the problem allows to apply a direct Monte Carlo method to simulate the binary collision dynamics with a computational cost directly proportional to the number of particles $N$ of the system, as described in~\cite{pareschi2013interacting}. The computational efficiency is important for parameter identification purposes since the segmentation masks  to optimize corresponds to the large time distribution of the system. Applications to biomedical images will be presented to test the performance of the proposed method. 
    
    In more detail, the manuscript is organized as follows. In Section~\ref{sec2} we introduce the generalized Hegselmann-Krause model for image segmentation. The evolution of aggregate quantities are then obtained by means of a Boltzmann-type equation in Section~\ref{subsect:Boltzmann}. We derived a surrogate Fokker-Planck model for which the large time distribution can be efficiently computed and compared with the one of the Boltzmann-type model. In Section~\ref{sec4} we describe how to generate segmentation masks and in section~\ref{subsec4.1} the strategy to optimize the model parameters is discussed. In Section~\ref{subsec5.2} we apply the proposed segmentation pipeline to three different biomedical image datasets. Finally, in Section~\ref{subsec5.4} we propose a patch-based version of the method that we apply for the segmentation of Magnetic Resonance Images of the thigh muscles. Our numerical experiments show that the introduced segmentation method achieves good performances for the tested images. 
    
    \section{Modelling consensus dynamics}\label{sec2}
    The study of consensus formation has gained great interest in the field of opinion dynamics to understand the basic ingredients underlying the phenomena of choice formation in connected communities. Several mathematical models of consensus have been proposed in the form of systems of first order ordinary differential equations (ODEs) or binary algorithms describing the behaviour of a finite number of particles/agents. In this direction, we mention the pioneering works \cite{french56,degroot74} where simple agent-based models were introduced to observe the relative influence among individuals, see also \cite{CS77} for a stochastic version of the above information exchange processes. 
    
    In recent years, several differential models of consensus have been introduced to understand the underlying social forces of social phenomena. Without intending to review all the literature, we point the interested readers to some references for finite systems: in \cite{SWS00,Galam97} simple Ising spin models have been introduced to mimic the mechanisms of decision making in closed communities, in \cite{DNAW00} a binary scheme is adopted to measure the convergence towards consensus under interaction limitations. Furthermore, models incorporating leader-follower effect have been considered in \cite{NC10} and models of social interaction on networks have been studied in \cite{albipar2017, WD07}. Finally, in \cite{MT14} a microscopic modelling approach is considered to measure convergence towards consensus in the case of asymmetric interactions. For a review we mention \cite{castellano09}. We highlight how consensus-like dynamics may model heterogeneous phenomena, see e.g. flocking dynamics \cite{CS07,bertozzi06,SeungYeal2008} or economic interactions \cite{degond14,mezard}. 
    
    Beside microscopic particle-based models for consensus dynamics, in the limit of infinitely many agents, it is possible to cope with the evolution of distribution functions characterizing the aggregate trends of the interacting systems. These approaches are generally based on kinetic-type partial differential equations (PDEs) and are capable of linking microscopic forces to emerging features of the system. In this direction we mention \cite{AHP2015,APZ14,BT,DW,DWright,PPDT,T06} and the references therein. For macroscopic models of consensus dynamics we point the interested reader to \cite{CG19,PTTZ}.
    
    \subsection{The bounded confidence model}
    In the following, we consider a population of $N\ge 2$ particles characterized by an initial state $x_i(0) \in X \subseteq \mathbb{R}$. Each particle $i \in \{1,\dots,N\}$ modifies its state $x_i\in X$ through the interaction with the particle $j \in \{1,\dots,N\}$, $j\ne i$, only if $x_i$ is sufficiently close to $x_j$, i.e. $|x_i-x_j| \le \Delta$, being $\Delta >0$ a suitable threshold. This model stresses the homophily in learning processes and is known as Hegselmann-Krause (HK) model \cite{hegselmann2002opinion}, see \cite{lorenz07} for a survey. At the time-continuous level, the dynamics of the $i$th particle may be suitably defined as follows 
    \begin{equation}\label{eq:HK_1D}
       \dfrac{d}{dt} x_i = \alpha\sum_{j=1}^N \chi(|x_j-x_i|\leq \Delta) (x_j-x_i) \qquad i=1,\dots,N 
    \end{equation}
    where $\chi(\cdot)$ is a characteristic function and  defines the bounded confidence interaction scheme, and $\alpha>0$ is a suitable scaling constant measuring that contribution of each interaction. Generally, it is assumed $\alpha = 1/N$ such that, if $X  = [-1,1]$ and $\Delta =2$ the system converges towards the mean $\bar x = \frac{1}{N}\sum_{j=1}^N x_j(t)$. 
    
\begin{figure}
    \centering 
    \subfigure[$\Delta=1$]{\label{fig:0a}\includegraphics[width=0.325\textwidth]{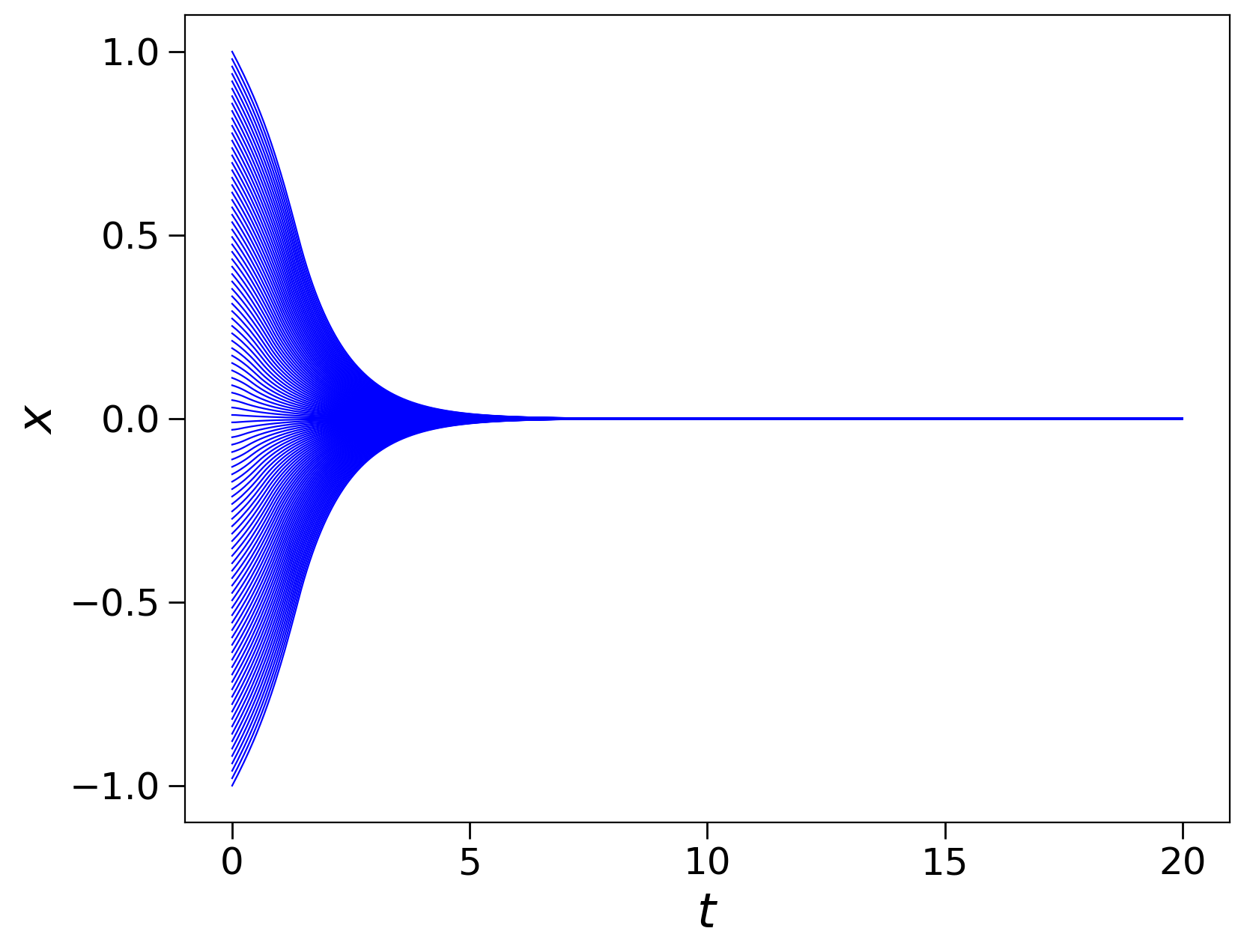}}
    \subfigure[$\Delta=0.5$]{\label{fig:0b}\includegraphics[width=0.325\textwidth]{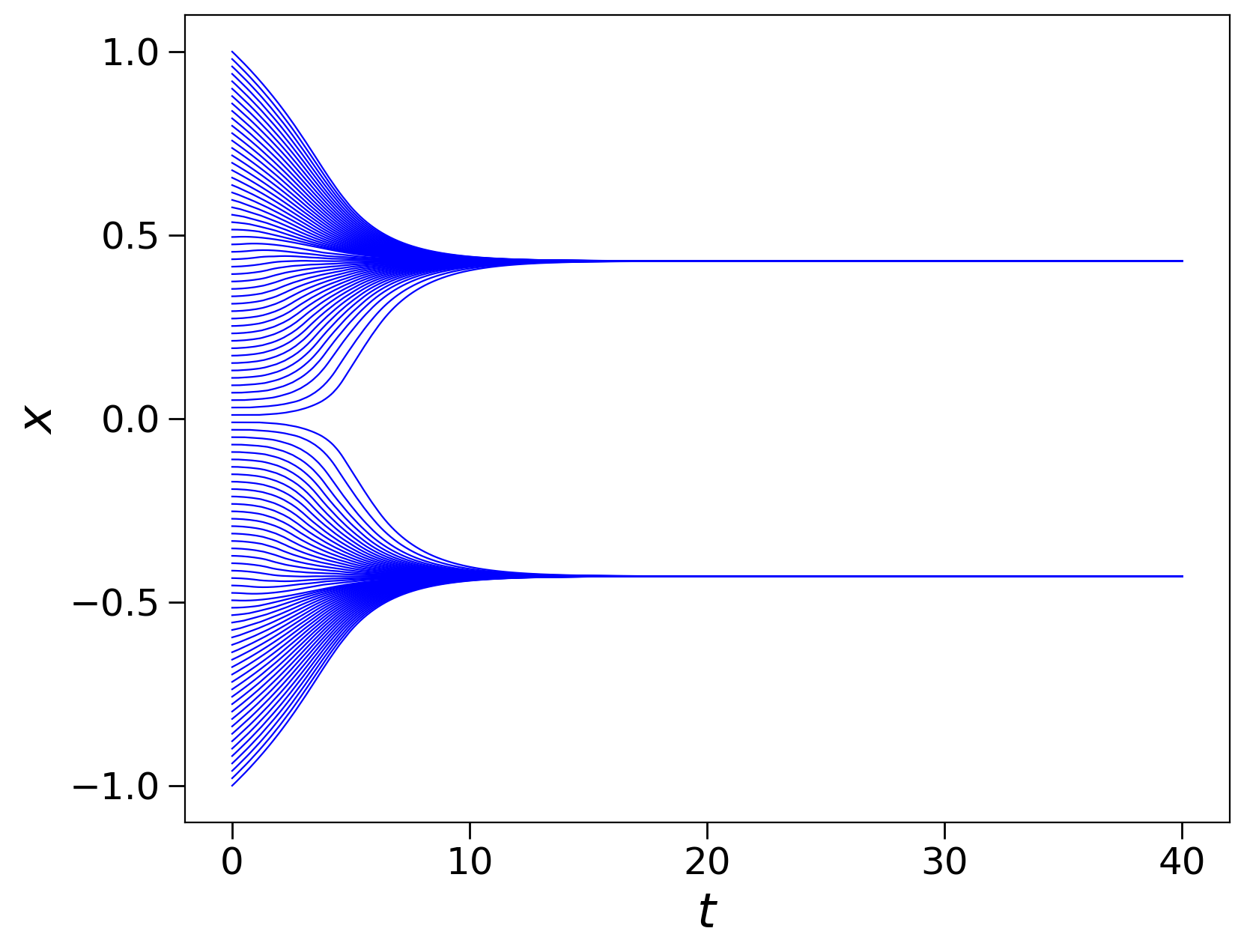}}
    \subfigure[$\Delta=0.3$]{\label{fig:0c}\includegraphics[width=0.325\textwidth]{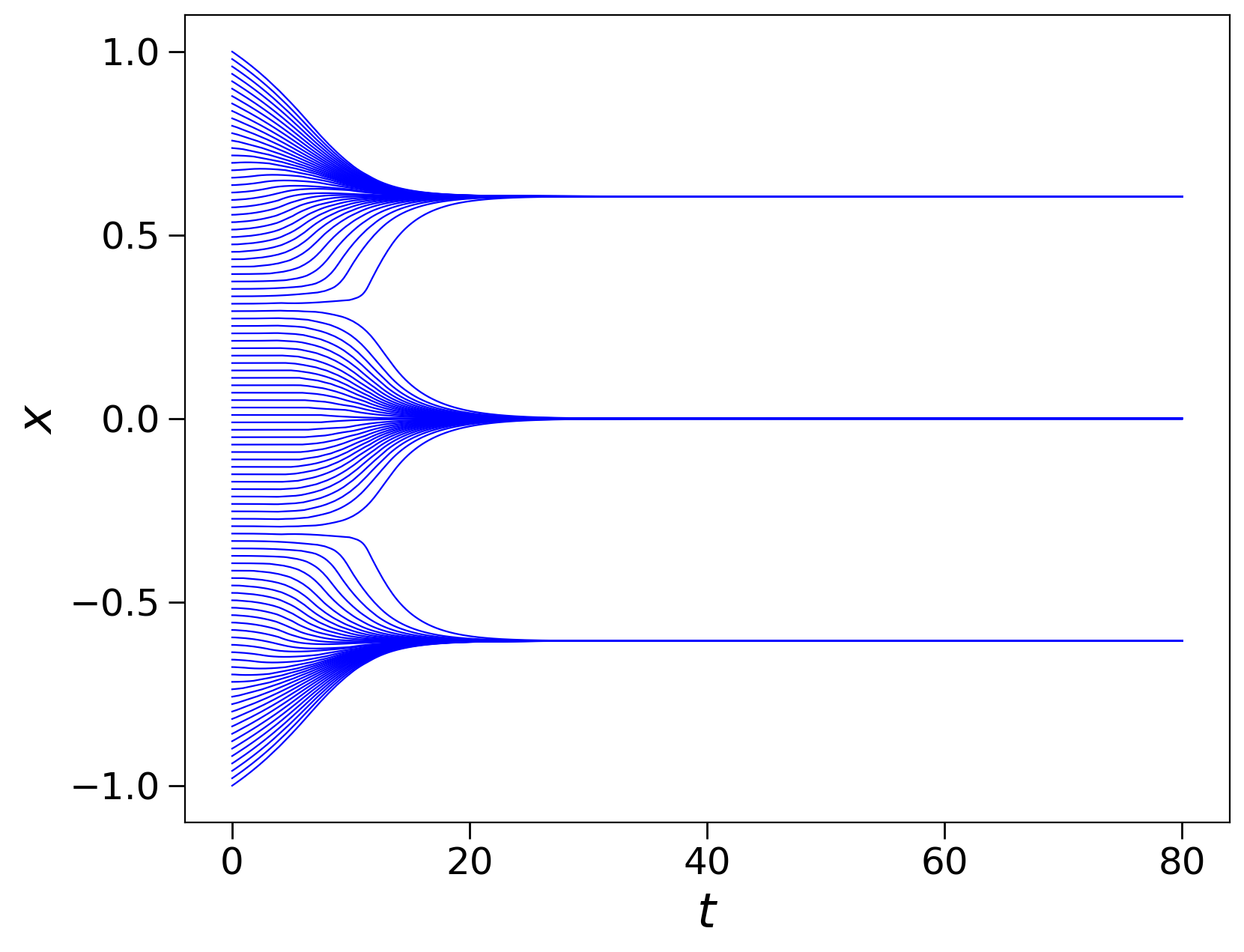}}
    \caption{Results of the Hegselmann-Krause bounded confidence model for three different values of the $\Delta$ threshold. At the initial time we selected $N=100$ particles equally spaced in $[-1,1]$.}
    \label{fig:0}
    \end{figure}

	Furthermore, since the interaction function $\chi(|x_j-x_i|\leq \Delta)\geq 0$, it has been proved that the HK model converges to a steady configuration where the initial states are grouped in a finite number of clusters, see~\cite{blondel2010continuous,canuto12,MT14}. In Figure~\ref{fig:0} we show the dynamics of a system of $N = 100$ particles according to the HK model for different values of the threshold $\Delta>0$. We may easily observe how multiple clusters appear for small values of $\Delta>0$.
     
    \subsection{Consensus models in segmentation problems}
    
	An application of the HK model to image segmentation problems has been proposed in~\cite{herty2020mean}. The key idea of this approach is to link  each particle $i\in \{1,\dots,N\}$ to a time-dependent position  $ \mathbf{x}_i = (x_i(y),y_i(t)) \in \mathbb{R}^{2} $ and a scalar quantity expressing a feature $c_i \in [0,1]$ and corresponding to a static characteristic of the system. In particular, in the following $c_i \in [0,1]$ expresses the gray level of the $i$th pixel. In this setting, the consensus process defined in \eqref{eq:HK_1D} depends also on the feature of the system and assumes the following form
    \begin{equation}
        \label{eq:HK_seg}
        \begin{aligned}
            & \dfrac{d}{dt} \mathbf{x}_i = {1 \over N} \sum _{j=1}^N P_{\Delta_1,\Delta_2}(\mathbf{x}_i, \mathbf{x}_j, c_i, c_j) (\mathbf{x}_{j}-\mathbf{x}_{i})  \\
            & \dfrac{d}{dt}c_{i}= 0 \quad i=1,\dots, N \\
        \end{aligned}
    \end{equation}
    where $\mathbf{x}_i(0) = \mathbf{x}_i^0$, $c_i(0) = c_i^0$ and where we introduced the interaction function 
    \begin{equation}\label{eq:HK_P}
        P_{\Delta_1,\Delta_2}(\mathbf{x}_i, \mathbf{x}_j, c_i, c_j) = \chi(|\mathbf{x}_j-\mathbf{x}_i|\leq \Delta_1) \chi(|c_j-c_i|\leq \Delta_2)
    \end{equation}
    and $\Delta_1 \geq 0$ and $\Delta_2 \geq 0$ are the confidence intervals of the position vectors and of the gray levels, respectively. In \eqref{eq:HK_P} the function $\chi(\cdot)$ is the characteristic function. The two introduced scalar and image-dependent quantities are particularly important to determine the optimal segmentation masks.  Indeed, they determine the confidence level under which the $i$th pixel tends to form a cluster through interactions with the whole set of pixels determining the image. 
 
    It is important to highlight that biomedical images are often affected by ambiguities due to several sources of uncertainty linked to both clinical factors and to possible bottlenecks in data acquisition processes \cite{barbano22,kendall17}. Among them it is possible to distinguish two major types of uncertainty, we may refer to the first as aleatoric uncertainty and is linked to stochasticities in the data collection process. In this case, we have to face reconstruction problems in which the image processing models suffer raw acquisition data.    On the other hand, the second kind of uncertainty is of epistemic-type and determines deviation of model parameters. In particular, in medical imaging MRI (Magnetic Resonance Imaging) scans may lead to ambiguous segmentation outputs \cite{Kwon00}. In this regard, the study of uncertainty quantification in image segmentation is a growing field to produce robust segmentation algorithms that are capable of avoiding erroneous results. 
    
    Therefore, in order to take into account aleatoric-type uncertainties affecting the correct feature of an image we consider a stochastic consensus model. In particular, we concentrate on the stochastic version of \eqref{eq:HK_seg} whose form is defined as follows 
    \begin{equation}
    \label{eq:HK_seg_D}
        \begin{aligned}
            & d \mathbf{x}_i = {1 \over N} \sum _{j=1}^N P_{\Delta_1,\Delta_2}(\mathbf{x}_i, \mathbf{x}_j, c_i, c_j) (\mathbf{x}_{j}-\mathbf{x}_{i})dt+ \sqrt{2\sigma^2D(c_i)}d\mathbf{W}_i  \\
            & \dfrac{d}{dt}c_i = 0, \quad i=1, \dots, N
        \end{aligned}
    \end{equation}
    where the interaction function is compatible with \eqref{eq:HK_P}, $\mathbf{x}_i(0) \in \mathbb R^2$, $c_i(0) \in [0,1]$ and $\mathbf{W}_i$'s are independent Wiener processes. Furthermore, since marked deviations can be expected far from the boundaries of the features' domain, we consider a non-uniform impact of the introduced aleatoric uncertainty. To this end, we consider a local diffusion function $D(c)$ such that $D(0) = D(1) = 0$. A possible form of this function is $D(c)= c(1-c)$, $c\in [0,1]$. In \eqref{eq:HK_seg_D} the diffusion is weighted by  the parameter $\sigma^2>0$. In Section~\ref{subsec5.3}, other choices  of the local diffusion function $D(c)$ will be considered.

%%==================================%%
%% FOKKER PLANCK EQUATION %%
%%==================================%%

\section{Kinetic models for image segmentation}
In~\cite{herty2020mean} the mean-field limit of the generalized HK model is studied and it is formally argued that for $N \rightarrow \infty$ a system of particles whose dynamics is   \eqref{eq:HK_seg} can be described in terms of the following nonlocal partial differential equation  
        \begin{equation}\label{eq:MF}
            \begin{split}
               & \partial_t f(\mathbf{x}, {c}, t) = \nabla_\mathbf{x} \cdot  \left[ \mathscr{B}[f]_{\Delta_1, \Delta_2}(\mathbf{x}, c,t)) f(\mathbf{x}, c, t)\right] \\
                & f(\mathbf{x}, \mathbf{c}, 0) = f_0(\mathbf{x}, \mathbf{c})
            \end{split}
        \end{equation}
        where $\mathscr{B}_{\Delta_1, \Delta_2}[f](t, \mathbf{x}, c)$ is the operator that describes the interaction dynamics and is defined as
        \begin{equation}\label{eq:B}
            \mathscr{B}_{\Delta_1, \Delta_2}[f](\mathbf{x}, c,t) =\int_0^1 \int_{\mathbb R^2} P_{\Delta_1,\Delta_2}(\mathbf{x}, \mathbf{x}_*, c, c_*) (\mathbf{x} - \mathbf{x}_*) f(\mathbf{x}_*, c_*, t)  d \mathbf{x}_* d c_*,       
        \end{equation}
       where  the interaction function $P_{\Delta_1,\Delta_2}(\mathbf{x}, \mathbf{x}_*, c, c_*)$ is defined as in~\eqref{eq:HK_P}. In the following, we argue how the same model can be obtained in suitable limits from binary interaction dynamics.  
        
\subsection{Boltzmann-type derivation}
\label{subsect:Boltzmann}

In order to define consensus dynamics from the point of view of kinetic theory we set up a consistent binary scheme defining the interactions between pixels. To this end, inspired by \cite{CFRT10,pareschi2013interacting}, we consider the dynamics defined in \eqref{eq:HK_seg_D} for two particles characterized by positions $\x_i, \x_j \in \mathbb R^2$ and features $c_i,c_j \in [0,1]$. Hence, we introduce a time discretization at the level of particles with time step $\epsilon>0$. Setting 
\[
\begin{split}
&\x:= \x_i(t), \quad \x_*:= \x_j(t), \quad \x^\prime = \x_i(t + \epsilon), \quad \x_*^\prime = \x_j(t + \epsilon),  \\
&c:= c_i(t), \quad c_*:=c_j(t),\quad c^\prime:= c_i(t+\epsilon), \quad c_*^\prime:=c_j(t+\epsilon),
\end{split}
\]
we may discretise the stochastic differential equation with Euler-Maruyama scheme with time step $\epsilon>0$ to obtain the following binary dynamics 
\begin{equation}
\label{eq:two_Bol}
    \begin{aligned} 
       \mathbf{x}' &= \mathbf{x} + \epsilon P_{\Delta_1, \Delta_2}(\mathbf{x}, \mathbf{x}_*, c, c_*) (\mathbf{x}_*-\mathbf{x}) + \sqrt{2\sigma^2 D(c)} \eta \\
       c' &= c \\
       \mathbf{x}_*' &= \mathbf{x}_* +\epsilon P_{\Delta_1, \Delta_2}(\mathbf{x}_*, \mathbf{x}, c_*, c) (\mathbf{x}-\mathbf{x}_*) + \sqrt{2\sigma^2 D(c_*)} \eta\\
       c_*' &= c_*
    \end{aligned}
\end{equation}
where $\eta = (\eta_i,\eta_j) $ is a centered 2D Gaussian random variable such that
\begin{equation}\label{eq:B_scale}
\begin{split}
        \langle \eta \rangle = (0,0) \qquad  \langle \eta_i \eta_j \rangle  =  \Sigma_{ij}, \qquad i,j = 1,2 
\end{split}
\end{equation}
where $\left\langle \cdot \right\rangle$ denotes the integration with respect to the distribution of $\eta$ and $\Sigma = (\Sigma_{ij})_{i,j=1,2}$ is a diagonal matrix with unitary diagonal components. 
A first important consequence of the binary scheme \eqref{eq:two_Bol} with $D\equiv 0$ is that the support of the positions can not increase. Indeed, since $P_{\Delta_1,\Delta_2} \in [0,1]$ and $\epsilon\in (0,1)$ we have
\[
\begin{split}
|\x| &= |(1-\epsilon P_{\Delta_1,\Delta_2})\x + \epsilon P_{\Delta_1,\Delta_2} \x_*| 
\le (1-\epsilon P_{\Delta_1,\Delta_2}) |\x| + \epsilon P_{\Delta_1,\Delta_2} |\x_*| \\
&\le \max\{|\x|,|\x_*|\}.
\end{split}
\]
Furthermore, we can observe that, as in \cite{T06}, the binary interactions \eqref{eq:two_Bol} are such that
\begin{equation}
\label{eq:bin_m}
\begin{split}
\left\langle \x^\prime + \x_*^\prime \right\rangle &= \x + \x_* + \epsilon \left(P_{\Delta_1,\Delta_2}(\x,\x_*,c,c_*) - P_{\Delta_1,\Delta_2}(\x_*,\x,c_*,c)  \right)(\x_*-\x)\\
&= \x + \x_*,
\end{split}\end{equation}
meaning that, since the interaction function $P_{\Delta_1,\Delta_2}$ is symmetric, the mean is conserved on average in a single binary interaction. 

In general, for the introduced interaction function~ \eqref{eq:HK_P} the energy is not conserved. Indeed, at leading order for $\epsilon>0$ small enough we have
\begin{equation}
\label{eq:bin_m2}
|\left\langle \x^\prime \right\rangle|^2 + |\left\langle \x_*^\prime \right\rangle|^2 = |\x|^2 + |\x_*|^2 - 2\epsilon P_{\Delta_1,\Delta_2}|\x-\x_*|^2+o(\epsilon)
\end{equation}
Therefore, at leading order the energy of expected position is dissipated after collision being $P_{\Delta_1,\Delta_2} \geq 0$. A particularly simple case may be obtained by considering $P_{\Delta_1,\Delta_2}$ constant. In this case, from~\eqref{eq:bin_m} we deduce that the mean position is conserved in each binary interaction, whereas if $\sigma^2 = 0$, the mean energy is dissipated.

Let us introduce the distribution function $f = f(\x,c,t): \mathbb R^2 \times [0,1]\times \mathbb R_+ \rightarrow \mathbb R_+$ such that $f(\x,c,t)d\x dc$ is the fraction of particles which at time $t \ge 0$ are represented by their position in $\x \in \mathbb R^2$ and feature $c \in [0,1]$.
The evolution of $f$ undergoing binary interactions  \eqref{eq:two_Bol} can be described in the following Boltzmann-type kinetic equation
\begin{equation}
\label{eq:bol_strong}
\begin{split}
&\partial_t f(\x,c,t)\\
&\quad = \left\langle \int_0^1 \int_{\mathbb R^2} \dfrac{1}{{}^\prime J} (f({}^\prime \x,{}^\prime c,t) f({}^\prime\x_*,{}^\prime c_*,t)-f(\x,c,t)f(\x_*,c_*,t))d\x_* dc_*\right\rangle,
\end{split}
\end{equation}
where $({}^\prime \x,{}^\prime \x_*)$ are the pre-interaction positions which generate the postinteraction positions $(\x,\x_*)$ according to the interaction rule \eqref{eq:two_Bol}. Similarly, we denoted with $({}^\prime c,{}^\prime c_*)$ the preinteraction features generating the postinteraction features $(c,c_*)$. Anyway, following the scheme \eqref{eq:two_Bol}, the dynamics do not prescribe evolution of the features. Finally, we denoted by ${}^\prime J$ the Jacobian of the transformation $({}^\prime \x,{}^\prime \x_*)\rightarrow (\x,\x_*)$. 

Equation \eqref{eq:bol_strong} may be recast in weak form as follows
\begin{equation}\label{eq:Boltz}
\begin{split}
  & \dfrac{d}{dt} \int_{[0,1]}\int_{\mathbb R^2} \phi(\mathbf{x},c) f(\mathbf{x}, c,t) d\mathbf{x} dc\\ 
   &\quad= \int_{[0,1]^2}\int_{\mathbb R^4} \left \langle \phi(\mathbf{x}', c') -\phi(\mathbf{x}, c)\right \rangle f(\mathbf{x}, c,t) f(\mathbf{x}_*, c_*,t) d\mathbf{x} dc d\mathbf{x}_* dc_* ,
\end{split}\end{equation}
where $\phi(\x,c):\mathbb R^2\times[0,1]\rightarrow \mathbb R$ is a test function. Choosing $\phi(\x,c) = 1$, we get that the total mass of $f$ is constant in time, meaning that the total number of particles/pixels are conserved. Choosing instead $\phi(\x,c)=\x$ we get
\[
\begin{split}
&\dfrac{d}{dt} \int_0^1 \int_{\mathbb R^2} \x f(\x,c,t)d\x dc \\
&\quad=\epsilon \int_{[0,1]^2} \int_{\mathbb R^4} (P_{\Delta_1,\Delta_2}(\x,\x_*,c,c_*) - P_{\Delta_1,\Delta_2}(\x_*,\x,c_*,c))(\x_*-\x)  \\
&\quad\quad\times  f(\x,c,t)f(\x_*,c_*,t) d\x d\x_* dc dc_*.
\end{split}\]
Therefore, the mean position $M = \int_0^1 \int_{\mathbb R^2}\x f(\x,c,t)d\x dc$ is conserved in time being $P_{\Delta_1,\Delta_2}$ symmetric. 

In the following, we introduce a suitable scaling under which we can obtain the nonlocal mean-field model \eqref{eq:MF} starting from the binary interaction scheme \eqref{eq:two_Bol}. The procedure is based on the quasi-invariant regime introduced in \cite{T06}. We introduce the new time scale $\tau = \epsilon t$, we scale the  distribution function $g(\x,c,\tau) = f(\x,c,\tau/\epsilon)$, and we introduce the following scaling for the variance 
\begin{equation}
\label{eq:scale_sigma}
\sigma^2 \rightarrow \epsilon \sigma^2.
\end{equation}
We observe that $\partial_\tau g = \frac{1}{\epsilon} \partial_t f$ and the equation satisfied by $g$ is
\begin{equation}
\label{eq:g}
\begin{split}
&\dfrac{d}{d\tau}\int_0^1 \int_{\mathbb R^2} \phi(\x,c)g(\x,c,\tau)d\x dc \\
&\quad= \dfrac{1}{\epsilon} \int_{[0,1]^2}\int_{\mathbb R^4} \left\langle \phi(\x^\prime,c^\prime)-\phi(\x,c) \right\rangle  g(\x,c,\tau)g(\x_*,c_*,\tau) d\x d\x_* dc dc_*.
\end{split}
\end{equation}
Hence, if $\epsilon\ll 1$ and if $\phi$ is sufficiently smooth then the difference $\left\langle \phi(\x^\prime,c^\prime)-\phi(\x,c) \right\rangle$ is small and can be expanded in Taylor series. We obtain
\begin{align}\label{eq:Tay_Boltz}
  \langle\phi(\x^\prime, c^\prime) -\phi(\x, c) \rangle= \langle\x^\prime-\x\rangle \cdot \nabla_\x \phi(\x, c)  \\
  + \dfrac{1}{2}\langle(\x^\prime-\x)^T H[\phi] (\x^\prime-\x)\rangle + R_\epsilon(\x,\x_*,c),
\end{align}
where $R_\epsilon(\x,\x_*,c)$ is the remainder term of the Taylor expansion and $H[\phi]$ is the Hessian matrix. Within the scaling \eqref{eq:scale_sigma} we highlight that, by construction, the remainder term $R_\epsilon$ depends in a multiplicative way on higher moments of the random variable $\eta$ su that $R_\epsilon/\epsilon\ll 1$ for $\epsilon\ll 1$. We point the interested reader to \cite{CPT,PTknowledge} for further details. We note that all the terms multiplied by $c'-c$ do not appear in expression~\eqref{eq:Tay_Boltz} since $c^\prime = c$.

Hence, by substituting \eqref{eq:Tay_Boltz}  into equation \eqref{eq:g} we have
\begin{equation}
\label{eq:exp_Boltz}
    \begin{split}
        &\dfrac{d}{d\tau} \int_0^1\int_{\mathbb R^2} \phi(\x,c)  g(\x, c,\tau) d\x dc \\ 
       & \quad= \dfrac{1}{\epsilon} \int_{[0,1]^2}\int_{\mathbb R^4} \langle\x^\prime - \x\rangle\cdot \nabla_\x \phi(\x,c) g(\x, c,t) g(\x_*, c_*,\tau) d\x d\x_* dc dc_* \\
       &\qquad + \dfrac{1}{2\epsilon} \int_{[0,1]^2}\int_{\mathbb R^4} \langle(\x^\prime - \x)^T H[\phi(\x,c)] (\x^\prime - \x)\rangle g(\x, c,\tau) g(\x_*, c_*,\tau) d\x dc d\x_* dc_* \\
       & \qquad+ \dfrac{1}{\epsilon}  \int_{[0,1]^2}\int_{\mathbb R^4} R_\epsilon(\x,\x_*,c) \phi(\x,c)  g(\x, c,\tau) g(\x_*, c_*,\tau) d\x d\x_* dc dc_* 
    \end{split}
\end{equation}

In the limit $\epsilon \rightarrow 0^+$ we formally obtain:
\begin{equation}
\label{eq:lim_Boltz}
    \begin{split}
        &\dfrac{d}{d\tau} \int_0^1\int_{\mathbb R^2} \phi(\x,c) g(\x, c,\tau) d\x dc \\ 
       & \quad= \int_{[0,1]^2} \int_{\mathbb R^4} P_{\Delta_1, \Delta_2}(\mathbf{x}, \mathbf{x}_*, c, c_*)(\x_*-\x)\cdot \nabla_\mathbf{x} \phi(\x, c)
       g(\x, c,t) g(\x_*, c_*,\tau) d\x dc d\x_* dc_* \\
       & \qquad+ \sigma^2\int_{[0,1]^2} \int_{\mathbb R^4}  D(c) \nabla_\mathbf{x}^2 \phi(\x, c)g(\x, c,\tau) g(\x_*, c_*,\tau) d\x d\x_* dc dc_*.
    \end{split}
\end{equation}
Next, we may integrate back by parts and restoring the original variables we get the following Fokker-Planck-type equation
\begin{equation}\label{eq:FP}
    \partial_t g(\x, c, \tau) = \nabla_\x \cdot  \left[ \mathscr B_{\Delta_1,\Delta_2}[g](\x,c,t) g(\x,c,\tau) + \sigma^2 D(c) \nabla_\x g(\x, c, \tau)\right],
\end{equation}
provided the following boundary condition is satisfied 
\[
\begin{split}
\mathscr B_{\Delta_1,\Delta_2}[g](\x,c,\tau) g(\x,c,\tau) + {\sigma^2 }D(c) \nabla_\x g(\x, c, \tau ) \Bigg|_{|\x|\rightarrow +\infty} = 0 \\
\end{split}\]
{We highlight how the analytical equilibrium distribution of \eqref{eq:FP} is not generally available unless $P_{\Delta_1,\Delta_2} \equiv 1$, see e.g. \cite{T06} for a related setting. }
In \eqref{eq:FP} the nonlocal operator $\mathscr{B}_{\Delta_1,\Delta_2}[\cdot]$ corresponds to the one defined in \eqref{eq:B}. We may observe how the obtained model is equivalent to the Fokker Planck equation~\eqref{eq:MF} in absence of the  diffusion coefficient, i.e. in the case  $\sigma^2=0$. 
Furthermore, in the case $P_{\Delta_1,\Delta_2} \equiv 1$ and in the absence of diffusion we can prove that the energy is dissipated in time, consistently with what we observed in Section~\ref{subsect:Boltzmann}.

\subsection{DSMC method for Boltzmann-type equations}\label{subsect:DSMC}

In this section, we introduce a direct simulation Monte Carlo (DSMC)  approach to Boltzmann-type equations. The numerical approximation of nonlinear Boltzmann-type models is a major task and has been deeply investigated in the recent decades, see e.g. \cite{albi2013binary,dimarco14,pareschi2001introduction,pareschi2013interacting}. The main issue of deterministic methods relies on the so-called curse of dimensionality affecting the approximation of the multidimensional integral of the collision operator. Furthermore, the preservation of relevant physical quantities is challenging at the deterministic level, making the schemes model dependent. On the other hand Monte Carlo methods for kinetic equations naturally employ the microscopic dynamics defining the binary scheme to satisfy the physical constraints and are much less sensitive to the curse of dimensionality. Among the most popular examples of Monte Carlo methods for the Boltzmann equation we mention the DSMC method of Nanbu \cite{nanbu}, Bird \cite{bird}. Rigorous results on the convergence of the methods have been provided in \cite{babovsky1989convergence}. Furthermore,  the computational cost of the method is $O(N)$, where $N$ is the number of the particles of the system, making this method very efficient. 

In the following, we describe the DSMC method based on the Nanbu-Babovsky scheme. This method consists in selecting particles by independent pairs and making them evolve at the same time according to the binary collision rules described in equation~\eqref{eq:two_Bol}. Let us consider a time interval $[0,T]$ and discretize it in $N_t$ intervals of size $\Delta t$. We introduce the stochastic rounding of a positive real number $x$ as 
\[
\textrm{Sround}(x) = 
\begin{cases}
\left\lfloor x \right\rfloor+1 & \textrm{with probability}\quad x - \lfloor x \rfloor \\
\left \lfloor x \right\rfloor     & \textrm{with probability } \quad 1 - x+\left \lfloor x\right\rfloor,
\end{cases}
\]
where $\left\lfloor x\right\rfloor$ denotes the integer part of $x$. 
        We sample the random variable $\eta$ from a 2D Gaussian distribution with mean equal to zero and diagonal covariance matrix. 
        
        \begin{algorithm}[h]
            \caption{Monte Carlo algorithm for Boltzmann equation \eqref{eq:Boltz}}\label{alg:Boltz}
            \begin{algorithmic}[1]
                \State Given $N$ particles $(\textbf{x}_n^0 , c_n^0)$, with $n = 1, \dots, N$ computed from the initial distribution $f_0(\textbf{x}, c)$;
                \For{$t=1$ \textbf{to}  $T$}
                \State set $n_p=\textrm{Sround}(N/2)$;
                \State sample $n_p$ pairs $(i,j)$ uniformly without repetition among all possible pairs of particles at time step $t$;
                \State for each pair $(i,j)$, sample $\eta$
                \State for each pair $(i,j)$, compute the data change:
                \begin{equation}
                    \begin{aligned}
                       \Delta\mathbf{x}_i^t &= \epsilon P_{\Delta_1, \Delta_2}(\mathbf{x}_i^t, \mathbf{x}_j^t, c_i^0, c_j^0) (\mathbf{x}_j^t-\mathbf{x}_i^t) + \sqrt{2\sigma^2D(c_i^0)} \boldsymbol{\eta} \\
                       \Delta\mathbf{x}_j^t &= \epsilon P_{\Delta_1, \Delta_2}(\mathbf{x}_j^t, \mathbf{x}_i^t, c_j^0, c_i^0) (\mathbf{x}_i^t-\mathbf{x}_j^t) + \sqrt{2\sigma^2D(c_j^0)} \boldsymbol{\eta}
                    \end{aligned}
                \end{equation}
                compute
                    \begin{equation}
                       \mathbf{x}_{i,j}^{t+1} = \mathbf{x}_{i,j}^t + \Delta \mathbf{x}_{i,j}^t % \qquad
                       %x_j^{t+1} = x_j^t + \Delta x_j
                    \end{equation}
                \EndFor
            \end{algorithmic}
        \end{algorithm}
        The kinetic distribution, as well as its moments, is then recovered from the empirical density function 
        \[
        f_N(\x,c,t) = \dfrac{1}{N} \sum_{i=1}^N \delta(\x-\x_i(t))\otimes\delta(c-c_i(t)),
        \]
        where $\delta(\cdot)$ is the Dirac delta function. Hence, for any test function $\phi$ we denote the moments of the distribution $f$ by
        \[
        (\phi,f)(t) = \int_0^1\int_{\mathbb R^2} \phi(\x,c)f(\x,c,t)d\x dc,
        \]
        we have
        \[
        (\phi,f_N)(t) = \dfrac{1}{N} \sum_{i=1}^N \phi(\x_i,c_i).
        \]
        In the following, we will evaluate the approximation of the empirical density $f_N$ obtained by 
        \begin{equation}
        	\label{eq:approxS}
        f_{N,\Delta \x, \Delta c}(\x,c,t) = \dfrac{1}{N}\sum_{i=1}^N S_{\Delta \x}(\x-\x_i(t))\otimes S_{\Delta c}(c-c_i(t)),
        \end{equation}
        with $S_{\Delta \x}(\x)$, $S_{\Delta c}(c) $ suitable mollifications of the indicator function. In the simplest setting, if we consider the indicator function it would lead to the standard histogram. 
          
\subsection{Numerical examples for the Hegselmann-Krause dynamics}

In this section we provide numerical evidence of the consistency of the Boltzmann-type approach with respect to the one introduced in \eqref{eq:FP} corresponding, in the zero-diffusion limit to the model \eqref{eq:MF}.   In particular,  {since the analytical equilibrium distribution of \eqref{eq:FP} is not available we consider a} numerical solution of the Fokker Planck equation \eqref{eq:FP} {which preserves the equilibrium structure of the equation}, see \cite{pareschi2018structure}. This class of {structure preserving} schemes are capable of approximating with arbitrary order of accuracy the steady state of a Fokker-Planck-type model and preserve important physical properties like positivity and entropy dissipation. {Hence, we compare the numerical solution to \eqref{eq:FP} with the one of } the Boltzmann model, obtained through the Algorithm \ref{alg:Boltz}. 
       
        We consider in particular $N={10}^5$ particles uniformly distributed in the domain $[-1,1]^2\times[0,1] \subset \mathbb R^2 \times [0,1]$ 
         \begin{equation}
         \label{eq:f0}
            f(\x,c,0) =
            \begin{cases}
            \beta & (\x,c) \in [-1,1]^2 \times [0,1]\\
            0       & \textrm{elsewhere}
            \end{cases} 
        \end{equation}
        with $\beta>0$ a normalization constant such that $\int_{0}^1 \int_{\mathbb R^2}f(\x,c,0)d\x dc = 1$. Hence, we compare solution of the Boltzmann-type model \eqref{eq:bol_strong} under the scalings $\epsilon={10}^{-1}$ and $\epsilon={10}^{-2}$ with the solution of the Fokker-Planck model \eqref{eq:FP}. We consider the local diffusion function $D(c)$ of the form
$ D(c)=c(1-c)$, $c \in [0,1]$.

        Let us introduce the spatial grid $x_i \in [-L,L]$, $y_j \in [-L,L]$, $c_k \in [0,1]$ such that $x_{i+1}-x_i = \Delta x$, $y_{j+1}-y_j = \Delta y $, $c_{k+1}-c_k = \Delta c$. We further assume $\Delta x = \Delta y$ such that $L= N_x\Delta x = N_y\Delta y$ and $N_x = N_y = 61$, the discretization of the features' interval $[0,1]$ is instead computed with $N_c = 31$ gridpoints.  We use a second order semi-implicit method for the time integration of \eqref{eq:FP}, we refer to~\cite{pareschi2018structure} for a detailed description of these methods.  

        \begin{figure}
        \centering     %%% not \center
        \subfigure[]{\label{fig:1a}\includegraphics[scale = 0.3]
        {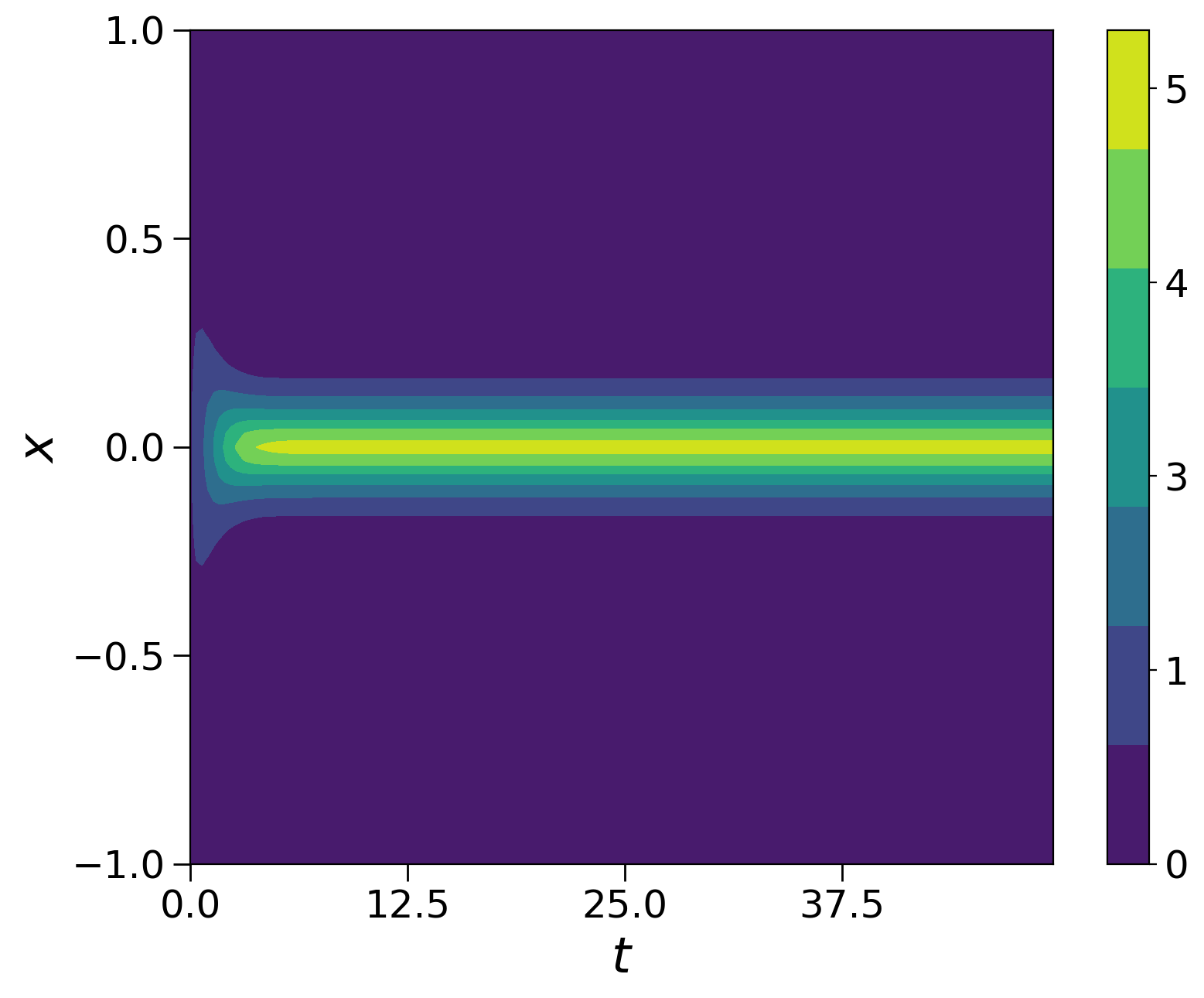}}
        \subfigure[]{\label{fig:1b}\includegraphics[scale = 0.3]
        {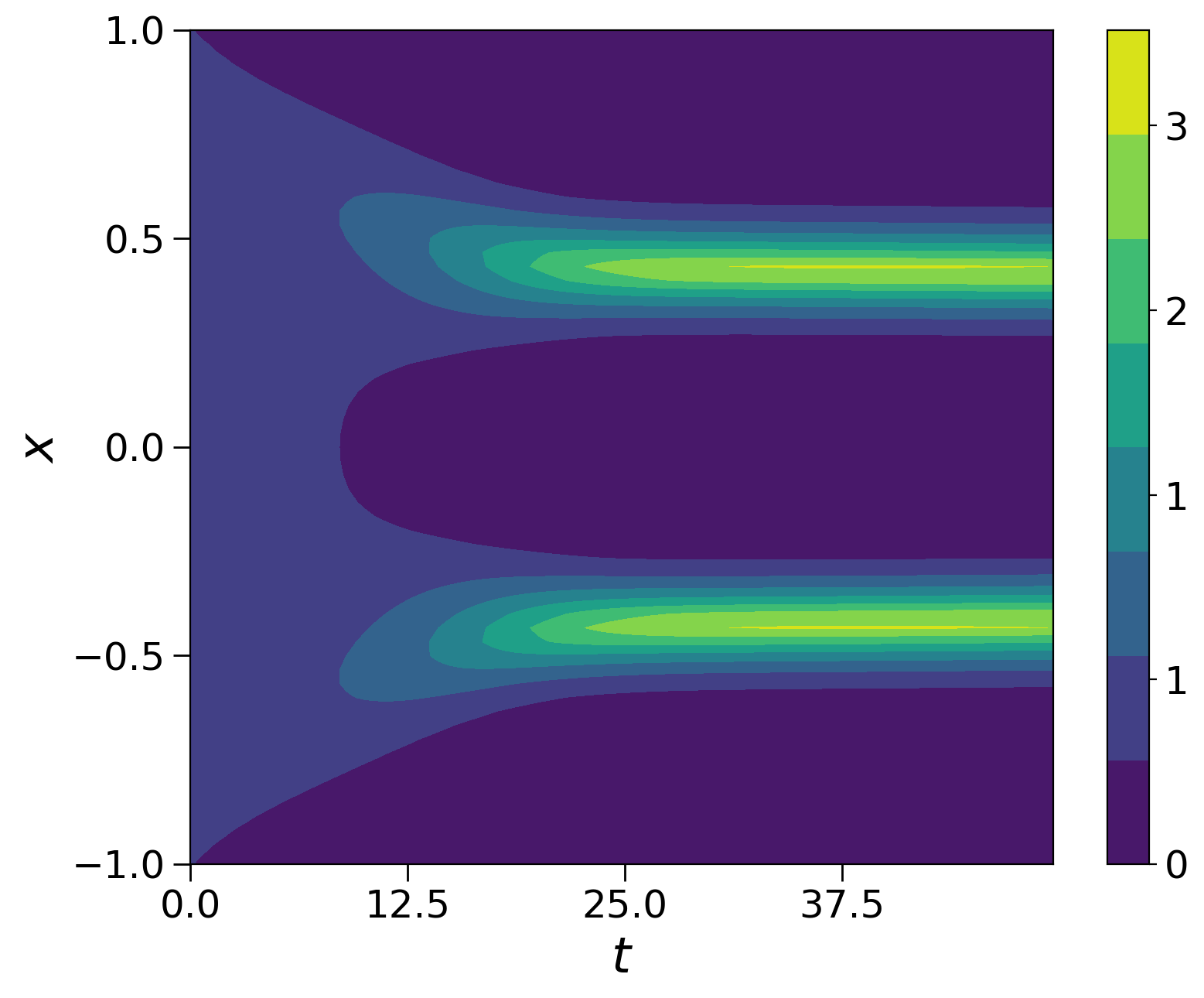}}
        \caption{Transient solutions of the Fokker-Planck equation \eqref{eq:FP} approximated through a semi-implicit SP scheme over $[0,T]$, $T=50$, with time step $\Delta t = 3 \cdot 10^{-1}$. In the left panel we consider $\Delta_1=2$, $\Delta_2=1$ and $\sigma^2= 5 \cdot 10^{-2}$ while in the right panel $\Delta_1=0.5$, $\Delta_2=1$ and $\sigma^2= 10^{-2}$. The initial distribution is \eqref{eq:f0}.}
        \label{fig:1}
        \end{figure}
        
        \begin{figure}
        \centering 
        \subfigure[]{\label{fig:2a}\includegraphics[width=\textwidth]{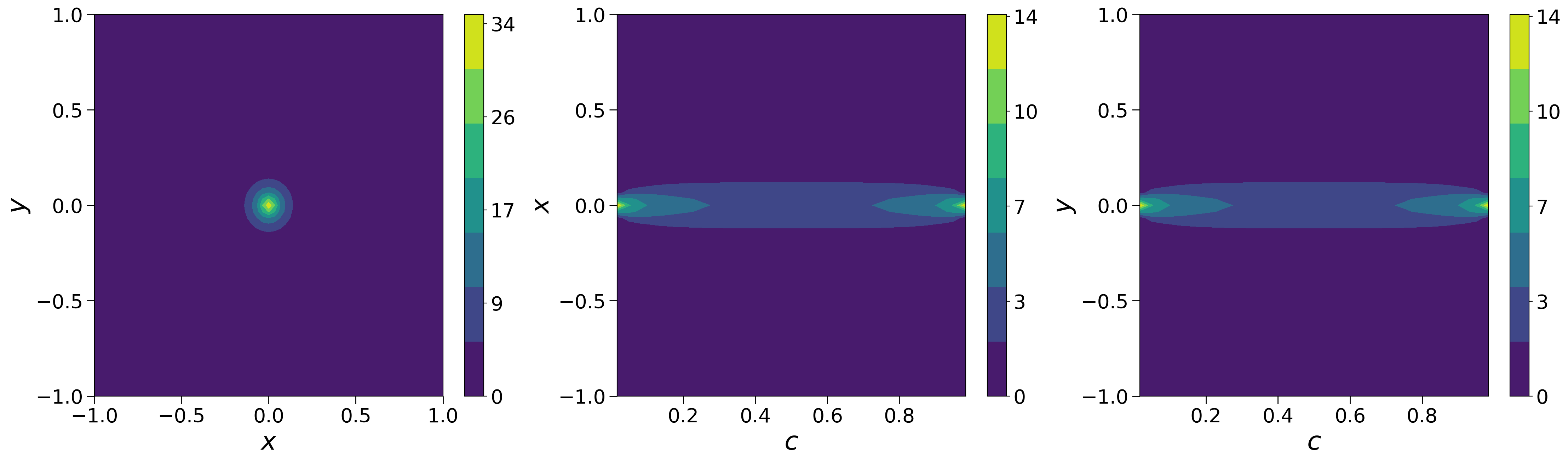}}
        \subfigure[]{\label{fig:2b}\includegraphics[width=\textwidth]{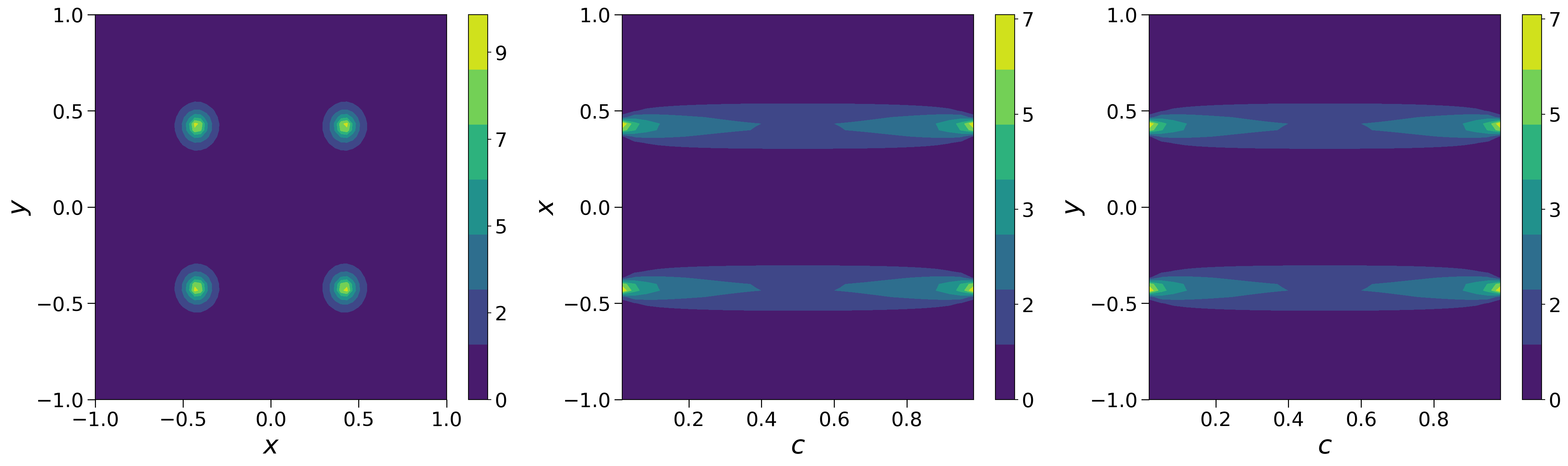}}
        \caption{Asymptotic solutions of the Fokker-Plank equation numerically computed with the SP scheme at the final time $T=50$. The left column represents the $xy$-projections, the middle column the $xc$-projections and the right column the $yc$-projections. In the top row we use $\Delta_1=2$, $\Delta_2=1$ and $\sigma^2= 5 \cdot 10^{-2}$ while in the bottom row $\Delta_1=0.5$, $\Delta_2=1$ and $\sigma^2= 10^{-2}$.}
        \label{fig:2}
        \end{figure}
        
        In Figure~\ref{fig:1} we show the transient behaviour of the Fokker-Planck solutions obtained with the semi-implicit SP scheme up in the time interval $[0,50]$ with time step $\Delta t = 3 \cdot 10^{-1}$ and we fix $\sigma^2 = 5 \cdot 10^{-2}$ for the left panel and $\sigma^2 = 10^{-2}$ for the right panel. We represent the projection of the kinetic density along the $x$-axis computed as $f_{yc}(x,t) = \int_{-L}^{L} \int_{0}^{1} f(x,y,c,t) dc dy$. In the left panel we fix the confidence coefficients $\Delta_1=2$ and $\Delta_2=1$ and in the right panel we choose $\Delta_1=0.5$ and $\Delta_2=1$. We may observe how the qualitative behavior of the solution dramatically changes since multiple clusters appear for large times, see \cite{PTTZ}.       
        
        Similarly, in Figure \ref{fig:2} we show the projections on the $xy$, $xc$ and $yc$ planes of the numerical solution of the introduced FP model at time $T = 50$. The projections are computed as $f_{c} = \int_{0}^{1} f(\x,c,T) dc$, $f_y = \int_{-L}^{L} f(\x,c,T) dy$, $f_x = \int_{-L}^{L} f(\x,c,T) dx$. 
        We may observe how, for large times, the kinetic density displays multiple clusters both the space variable $\x$ and in the features' variable $c$. The parameters $\Delta_1$, $\Delta_2$ and $\sigma^2$ are the same used in~\ref{fig:1a} and~\ref{fig:1b} for the top and bottom panels of Figure~\ref{fig:2} respectively.

        \begin{figure}
        \centering 
        \subfigure[]{\label{fig:3a}\includegraphics[scale = 0.3]
        {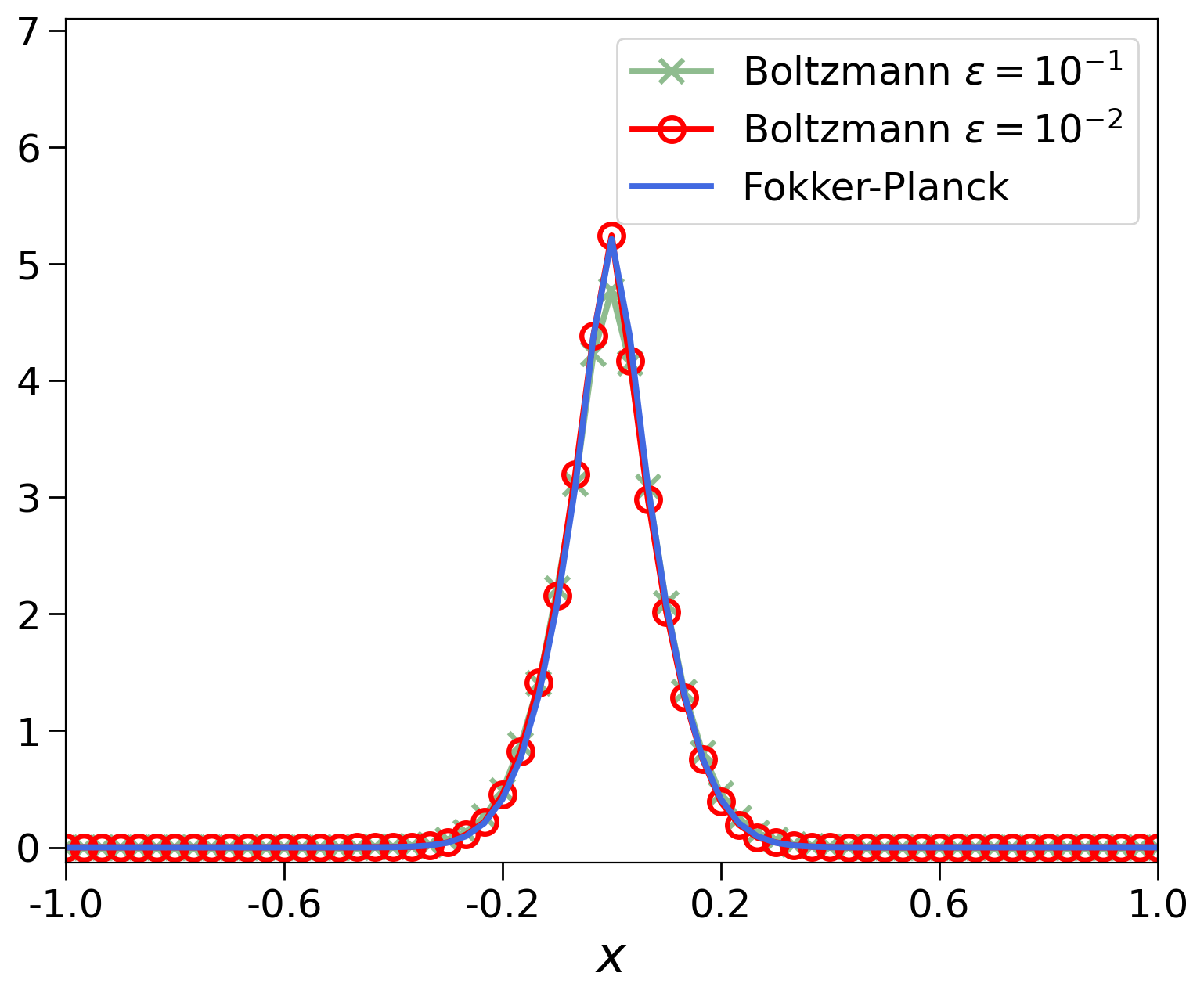}}
        \subfigure[]{\label{fig:3b}\includegraphics[scale = 0.3]
        {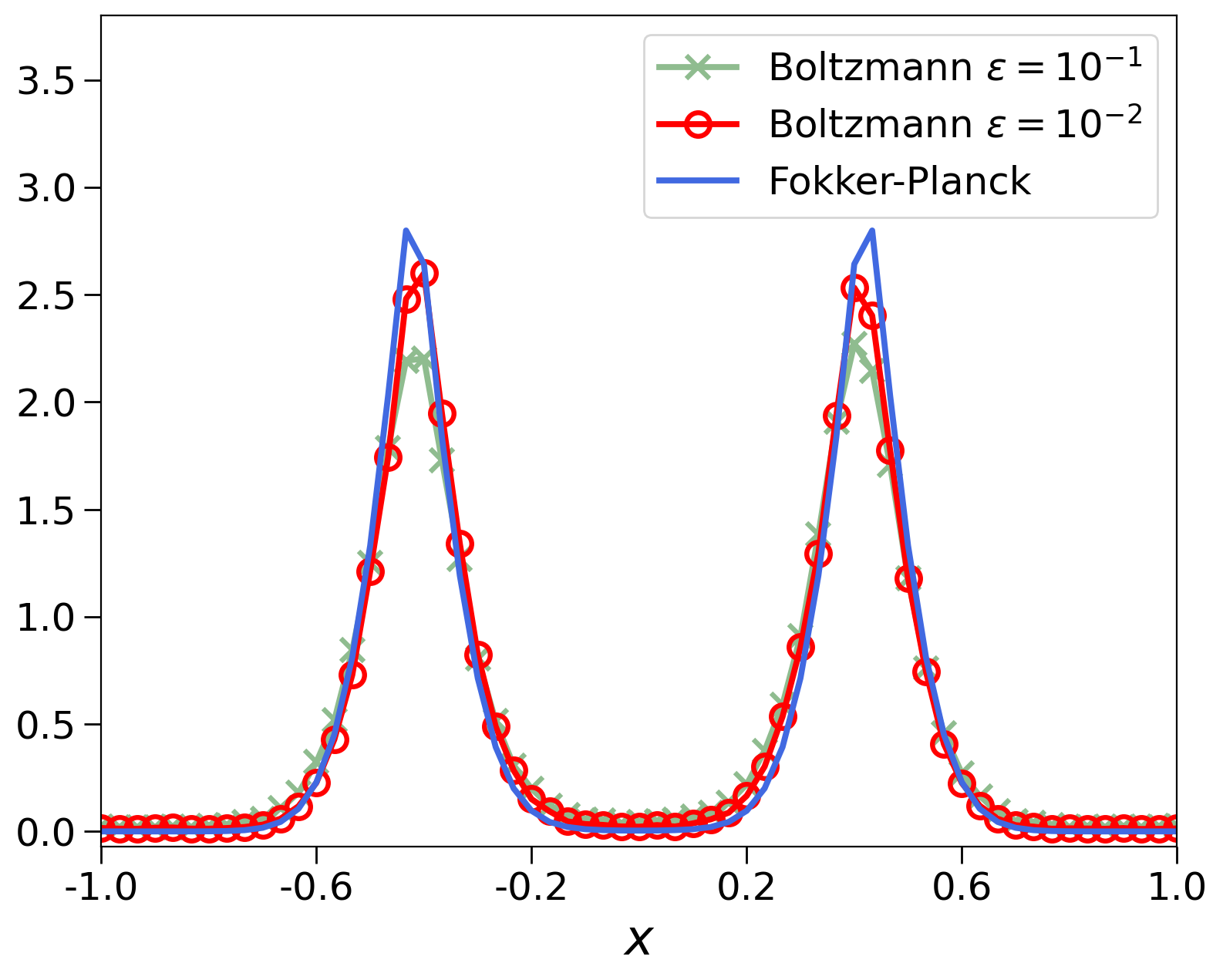}}
        \caption{Comparison between the numeric solution of the Fokker-Planck equation computed by the SP scheme (in blue) with final distribution provided by the MC algorithm for the Boltzmann-type equation with two different values of the parameter $\epsilon$ (in red and green). Both the panels represent the $x$-projections of the asymptotic distributions computed for $T=50$. In the left panel we use $\Delta_1=2$, $\Delta_2=1$ and $\sigma^2= 5 \cdot 10^{-2}$ while in the right panel $\Delta_1=0.5$, $\Delta_2=1$ and $\sigma^2= 10^{-2}$. The green distribution is computed with $\epsilon=10^{-1}$ and the red one with $\epsilon=10^{-2}$.}
        \label{fig:3}
        \end{figure}
        
        In Figure~\ref{fig:3} we show the x-axis projection of the Fokker-Planck solution provided by the SP scheme and the solutions of the Boltzmann equation obtained with Algorithm \ref{alg:Boltz} with $\epsilon = 10^{-2},10^{-1}$. We can also observe that, as expected, the solution of the Boltzmann model is a good approximation of the solution of the Fokker-Planck model for small values of the parameter $\epsilon>0$. 
        
        %%==================================%%
        %% SEGMENTATION %%
        %%==================================%%
        
        \section{Application to biomedical images}\label{sec4}
        In this section we focus on segmentation problems for medical images. In particular, we concentrate on the segmentation of images of cell nuclei, of brain tumour  and on the recognition of thigh muscles. In all the aforementioned cases, we apply Algorithm~\ref{alg:Boltz} to generate the segmentation masks. 

        The procedure to generate segmentation masks can be summarized as follows:
        \begin{enumerate}
        \item[$i)$] The pixels of the 2D image are interpreted as uniformly spaced particles, each characterized by the spatial position $(x_i,y_i)$ and with static feature $c_i$  defined are the gray level of the corresponding pixel. Hence, the initial distribution is reconstructed through \eqref{eq:approxS}.  In all the applications of this work we linearly scale the initial position of the particles on a reference domain $[-1,1]\times[-1,1]$ and the initial values of the features to the $[0,1]$ range.
        \item[$ii)$] We numerically determine the large time solution of the Boltzmann-type model in \eqref{eq:bol_strong} by means of the DSMC Algorithm~\ref{alg:Boltz}.  In this way, particles tend to aggregate in a number of finite clusters based on the Euclidean distance and on the difference between the gray levels of the pixels quantifying the features. 
        \item[$iii)$]  The segmentation masks are generated by computing the mean value of the features of pixels in the same cluster. We assign to these values to the initial position of the pixels. Using this method, homogeneous regions with similar characteristics are created in the image and correspond to the segmentation mask. 
        \item[$iv)$] The obtained multi-level mask can be transformed into a binary mask by fixing a threshold $\tilde c$ such that if $c_i<\tilde c$ then $c_i=0$ and if $c_i\ge \tilde c$ then $c_i = 1$. 
        \end{enumerate}
        
\begin{figure}
        \begin{center}
        \includegraphics[width=\textwidth]{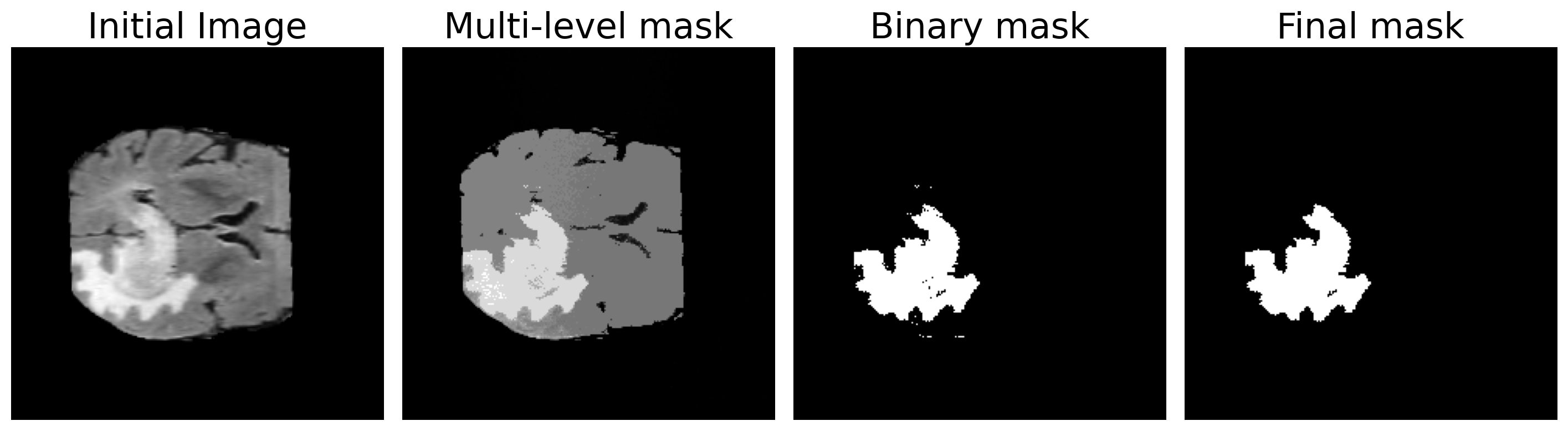}
        \caption{Summary of the segmentation process. The first panel shows the input image; the second panel displays the multi-level segmentation mask produced by DSMC Algorithm~\ref{alg:Boltz}; the third panel represents the mask after the binarization process; and the fourth panel shows the binary mask after the two morphological refinement steps.}
        \label{fig:steps}
        \end{center}
        \end{figure}

        Once the segmentation mask has been computed by this procedure, we apply two morphological refinement steps in order to remove those small regions misclassified as foreground parts and to fill possible small holes that were misidentified background parts. 
       Specifically, for the first step we label all the connected components of the foreground mask and remove the components whose number of pixels is below a fixed threshold. The second stage of morphological refinement works in the same way as the first, except that it is performed in a complementary manner on the background pixels. To implement these operations, we used the scikit-image python library~\cite{van2014scikit} which exploits the graph theory to find distinct objects of a binary image~\cite{fiorio1996two}. By using this approach, we were able to obtain more accurate segmentation masks by reducing some small imperfections of the segmentation process. The introduced segmentation procedure and the results of the two morphological steps are sketched in Figure~\ref{fig:steps}.

        %%==================================%%
        %% OPTIMIZATION %%
        %%==================================%%
        
        \subsection{Parameter identification}\label{subsec4.1}
        The purpose of this phase of the work is to choose the set of the optimal parameters $\Delta_1>0$, $\Delta_2>0$ and $\sigma^2>0$  for the segmentation problem based on the available data. 
        We perform the optimization procedure by systematically scanning the parameter search space through a random sampling process and measuring the segmentation performance to find the best combination of $\Delta_1$, $\Delta_2$ and $\sigma^2$.
        
        For image segmentation problems, the objective function quantifies the goodness of segmentation by measuring the distance or similarity between the ground truth segmentation mask and the evaluated mask.  Let us consider as objective function the Dice Similarity Coefficient ($DSC_{loss}$), which is defined as follows
        \begin{equation}
        \label{eq:dice}
            \begin{aligned}
            DSC_{loss} &= 1- DSC_{metric} \\
            &= 1 - {2  |S_{true}\cap S_{est}|\over |S_{true}| + |S_{est}|},
            \end{aligned}
        \end{equation}
        where $S_{true}$ is the true mask and $S_{est}$ is the estimated mask and the $DSC_{loss}$ is computed only on the foreground (white voxels). This index is null if there is a perfect overlap between the two masks and it is equal to one if the masks are completely disjoint \cite{yang2020parallelizable}. It follows that the best segmentation will minimize this index. {We chose the $DSC_{metric}$ in view of its computational efficiency and experimental explainability, making it suitable for optimizing the model parameters. We highlight how a variety of metrics may be considered to evaluate this task, see e.g. \cite{taha2015metrics} for a complete overview of segmentation metrics. Another relevant example is provided by Wasserstein distance which, however, suffers from high computational complexity and equivalent metrics have been introduced to maintain a good approximation \cite{auricchio_etal}.}  
        
        Hence, we solve the optimization problem 
        \begin{equation}
        \label{eq:min}
        \min_{\Delta_1,\Delta_2,\sigma^2>0} DSC_{loss}
        \end{equation}
        where the loss function is computed by looking at the numerical equilibrium of the model \eqref{eq:bol_strong} associated to a particular choice of parameters and obtained through Algorithm~\ref{alg:Boltz}. 

        To solve the optimization problem~\cref{eq:min}, we used the random sampling algorithm from the Hyperopt python package~\cite{bergstra2015hyperopt}. The key idea of this class of algorithms is to define a range of potential values for parameters of interest and randomly selecting a number of combinations to test. The performance of the objective function is evaluated for each point tested of the search space and the set of parameters that minimize the objective function represents the best configuration for solving the optimization problem. This technique can be useful for complex functions with non-linear, non-convex, or noisy shapes where gradient-based optimization may be difficult to apply~\cite{andradottir2006overview}. To determine the bounds for $\Delta_1$, $\Delta_2$ and $\sigma^2$, we performed the optimization on a sufficiently large search space and repeated the process by progressively limiting the search space to the most promising area.

It is crucial to note that the introduced optimization strategy currently relies on knowledge of ground truth segmentation. The primary aim of this process is the validation of the segmentation model, rather than focusing on practical applications in real-world scenarios where ground truth masks may be unavailable. Nevertheless, we are actively refining our approach to make it entirely data-oriented, enabling its implementation even in situations where ground truth data is not available.
        
    %%==================================%%
    %% APPLICATIONS %%
    %%==================================%%

         %%==================================%%
        %% 2D SEGMENTATION %%
        %%==================================%%
        \subsection{2D biomedical image segmentation}\label{subsec5.2}
        
        We apply the Monte Carlo Algorithm~\ref{alg:Boltz} and the optimization strategy described in section~\ref{subsec4.1} to 2D gray-scale biomedical image segmentation. We consider three different datasets:
        \begin{itemize}
        \item The \textit{HL60 cell nuclei dataset} is a public dataset collected for the Cell Tracking Challenge and available at \url{http://celltrackingchallenge.net/}. It consists of synthetic 2D time-lapse video sequences of fluorescent stained HL60 nuclei moving on a substrate, realized with the Fluorescence Microscopy (FM) technique. Each image is accompanied with the ground truth segmentation mask which identifies the cell nuclei. To test our 2D segmentation pipeline we use the first time frame of the sequences. 
        \item The \textit{brain tumor dataset} consists 3D in multi-parametric magnetic resonance images (MRI) of patients affected by glioblastoma or lower-grade glioma, publicly available in the context of the Brain Tumor Image Segmentation (BraTS) Challenge \url{http://medicaldecathlon.com/}. The acquisition sequences include $T_1$-weighted, post-Gadolinium contrast $T_1$-weighted, $T_2$-weighted and $T_2$ Fluid-Attenuated Inversion Recovery (FLAIR) volumes. 
        Three intra-tumoral structures were manually annotated by experienced radiologists, namely “tumor core”, “enhancing tumor” and “whole tumor”. We evaluate the performances of the MC algorithm for two different segmentation tasks: the  “tumor core” and the “whole tumor” annotations. 
        For the first task we use a single slice in the axial plane of the post-Gadolinium contrast $T_1$-weighted scans while for the second task we use a single slice in the axial plane of the $T_2$-weighted scans.
        \item The \textit{thigh muscles dataset} consists of 3D MRI scans of left and right thighs of healthy subjects and facioscapulohumeral dystrophy (FSHD) patients with muscle alterations. All the images were collected on a 3T MRI whole-body scanner (Skyra, Siemens Healthineers AG Erlangen, Germany) at the Mondino Foundation, Pavia Italy. Two acquisition protocols were performed for each subject: 3D six-point multi-echo gradient echo (GRE) sequence with interleaved echo sampling and a 2D multi-slice multi-echo spin echo (MESE) sequence. The dataset includes the ground truth segmentation mask of the 12 muscles of the thighs manually drawn by the radiologist team (we refer to~\cite{agosti2021deep} for a complete description of the acquisition settings and the annotation procedure). To test the segmentation pipeline we consider a single slice in the axial plane of the GRE scans of a healthy and an FSHD patient.
        \end{itemize}

        We optimize the values of the parameters $\Delta_1$, $\Delta_2$ and $\sigma^2$ by solving the minimization problem \eqref{eq:min}. For each dataset and segmentation task, we select the combination of the parameters that minimize the $DSC_{loss}$. The search space is defined by introducing additional constraints on the choice of parameters, in particular we consider $\Delta x\le\Delta_1\le 0.7$, $0.05\le\Delta_2\le 0.3$. The values of the parameter $\sigma^2$ are determined by sampling the data from a log-uniform distribution with support $[e^{-5},1]$. 
        The DSMC Algorithm~\ref{alg:Boltz} has been implemented in the quasi-invariant scaling with $\epsilon = 10^{-2}$ with final time $ T_{max} = 2000$.  
        
\begin{figure}
        \centering 
        \subfigure[]{\label{fig:4a}\includegraphics[width=\textwidth]{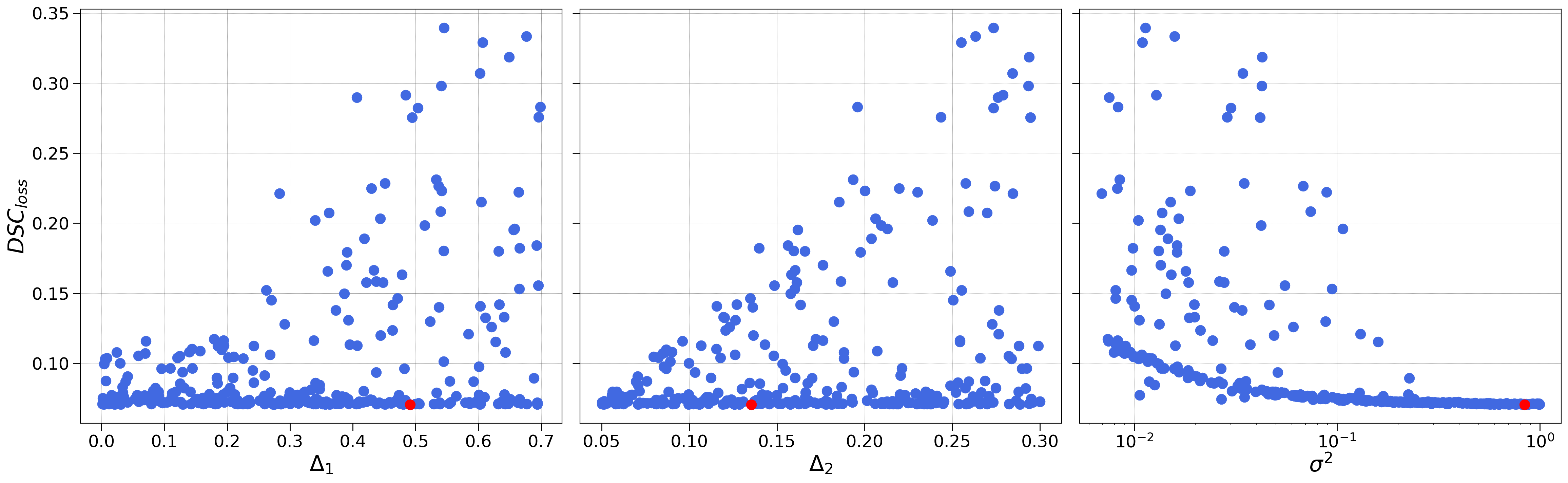}}
        \subfigure[]{\label{fig:4b}\includegraphics[width=\textwidth]{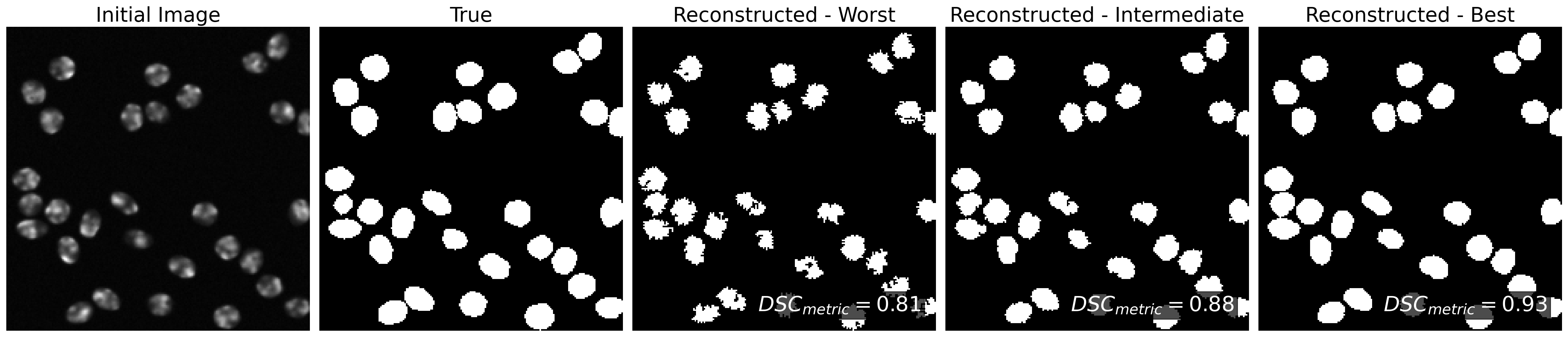}}
        \caption{ {Results of the optimization process for the \textit{HL60 cell nuclei dataset} as a function of the three parameters $\Delta_1$, $\Delta_2$ and $\sigma^2$. In panel~\ref{fig:4a}, we use scatter plots to graphically represent the distribution of the loss ($DSC_{loss}$) values as a function of the $\Delta_1$, $\Delta_2$ and $\sigma^2$ values. Red dots highlight the best configuration. In panel~\ref{fig:4b}, from left to right, we show the original image, the ground truth segmentation mask, and the masks computed using the worst, intermediate, and best parameter sets, respectively. In the bottom right corner, the corresponding values of $DSC_{metric}$ are displayed. The results in panel \ref{fig:4a} were obtained without morphological refinements, whereas those in panel \ref{fig:4b} include them.}}
        \label{fig:4}
        \end{figure}

        We present in Figure~\ref{fig:4} the results of the optimization process for the \textit{HL60 cell nuclei dataset}. In particular in Figure~\ref{fig:4a}, we show the values of the $DSC_{loss}$ function as a function of $\Delta_1$, $\Delta_2$ and $\sigma^2$ parameters.
In Figure~\ref{fig:4b}, we present the segmentations obtained with the best, intermediate, and worst configurations of the parameters $\Delta_1$, $\Delta_2$, and $\sigma^2$. It is important to highlight that both visually and in terms of the $DSC_{metric}$, the segmentation accuracy achieved with the optimal parameter configuration is better than that of the other configurations.
       
       The best configurations of the $\Delta_1$, $\Delta_2$ and $\sigma^2$ parameters for each dataset and segmentation task are listed in Table~\ref{table:best}.
        \begin{table}[h]
        \renewcommand\arraystretch{1.3}
        \centering
        \label{table:best}
        \begin{tabular}{ l c c c} 
        \hline 
        \textbf{Dataset name} & $\boldsymbol{\Delta_1}$ & $\boldsymbol{\Delta_2}$ & $\boldsymbol{\sigma^2}$ \\ 
        \hline 
        HL60 cell nuclei & 0.49 & 0.14 & 0.84 \\ 
        Brain tumor - tumor core & 0.69 & 0.17 & 0.03 \\
        Brain tumor - whole tumor & 0.34 & 0.28 & 0.25 \\
        Thigh muscles - healthy & 0.55 & 0.13 & 0.02 \\
        Thigh muscles - FSHD & 0.57 & 0.17 & 0.43 \\
        \hline
        \end{tabular}
        \caption{Best configurations of the $\Delta_1$, $\Delta_2$ and $\sigma^2$ parameters given by the optimization process.}
        \end{table}
        
        Once the best configurations of the $\Delta_1$, $\Delta_2$ and $\sigma^2$ parameters have been established, we generate the segmentation masks for all the datasets and annotation tasks solving the Boltzmann model in \eqref{eq:bol_strong} with the optimal choice of parameters. At the end of the segmentation procedure, we apply the two morphological refinement steps as described in Section~\ref{sec4} to improve the quality of the segmentation masks.
        
\begin{figure}
        \begin{center}
        \includegraphics[width=\textwidth]{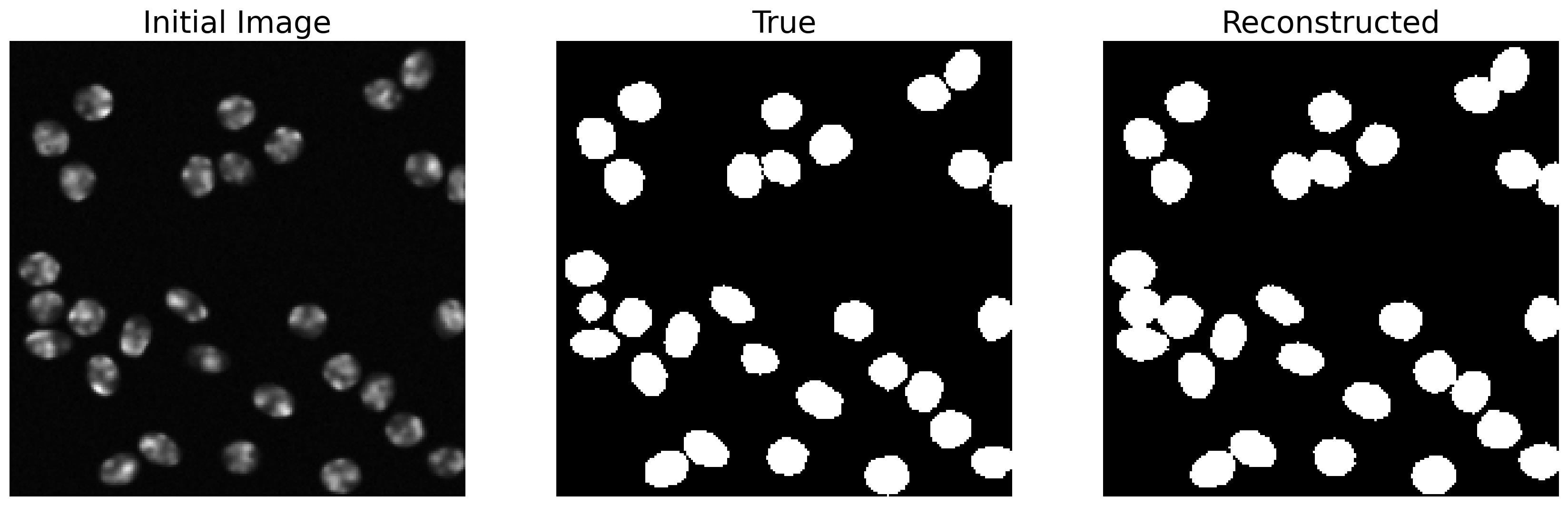}
        \caption{Results of the segmentation process for the \textit{HL60 cell nuclei dataset} obtained with the optimal set of $\Delta_1$, $\Delta_2$ and $\sigma^2$ parameters and with the morphological refinement steps. The three panels represent respectively from the left to the right the original image, the ground truth segmentation mask and the mask computed by the segmentation process.}
        \label{fig:5}
        \end{center}
        \end{figure}

        In Figure~\ref{fig:5} we represent the results of the segmentation process obtained for the \textit{HL60 cell nuclei dataset}. We can observe from the initial image that cells have regular borders but they are represented by a wide range of gray levels. By comparing the ground truth mask and the reconstructed one, we may observe that they are in good agreement. The method can provide good results in detecting the location of cells even for those regions whose gray level is close to the background. We evaluate the performance of the segmentation algorithm using the $DSC_{metric}$ coefficient defined in Equation~\eqref{eq:dice} which quantifies the overlap between the predicted and the true mask. The method reaches a $DSC_{metric}$ of $0.94$ on the image of the \textit{HL60 cell nuclei dataset} reported in Figure~\ref{fig:5}. 
Despite achieving the high $DSC_{metric}$ score, it is crucial to recognize that the method faces challenges in precisely distinguishing closely spaced cells. This limitation could potentially make the method less suitable for images characterized by high cell density. However, it is essential to underscore that the segmentations produced by our method can be further processed using supplementary approaches to address particular task requirements. For example, for improved cell separation, binary erosion methods can be applied to refine the shapes of the cell masks.
        
\begin{figure}
        \begin{center}
        \includegraphics[width=\textwidth]{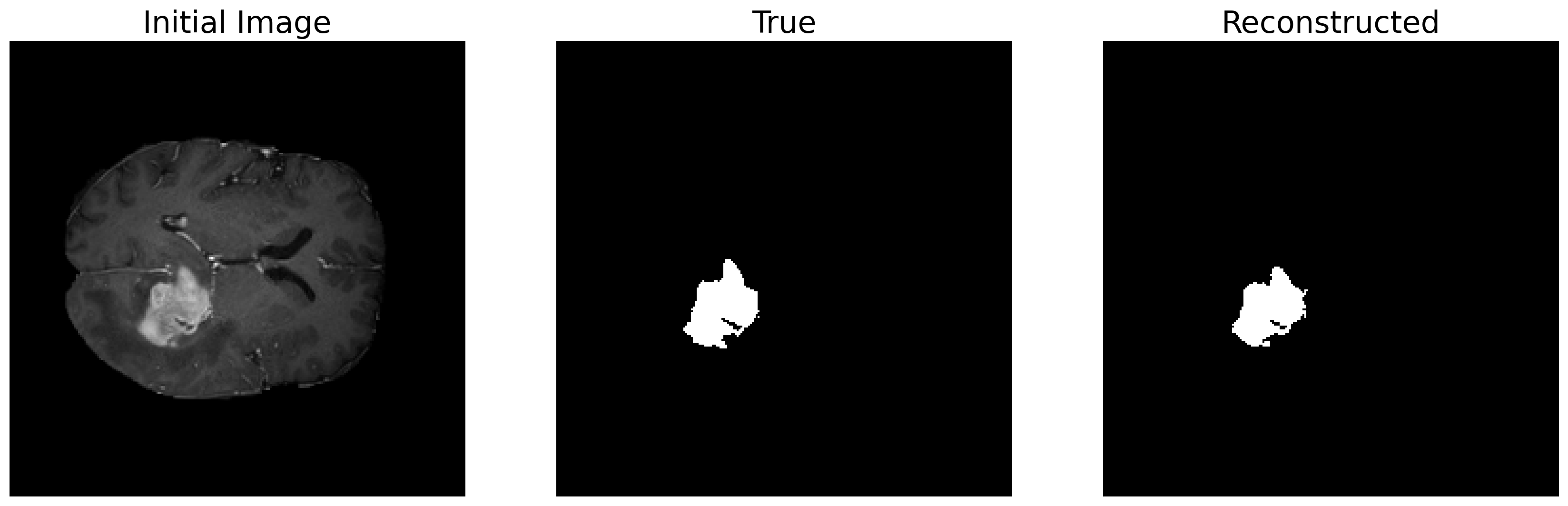}
        \caption{Segmentation results for the “tumor core” task of the \textit{brain tumor dataset}, obtained with the optimal set of $\Delta_1$, $\Delta_2$ and $\sigma^2$ parameters and with the morphological refinement steps.}
        \label{fig:6}
        \end{center}
        \end{figure}

        In Figure~\ref{fig:6} we test the performance of the method in segmenting the “tumor core” region of one image of the \textit{brain tumor dataset}. In this example, the region of interest is characterized by less regular edges, by the presence of cavities but by a quite homogeneous color. We can see from the Figure~\ref{fig:6} that the method identifies with a good precision the shape of the tumor and it preserves the small empty structures within the mass. In terms of the evaluation metric, the segmentation system achieves a $DSC_{metric}$ of $0.93$.
        
\begin{figure}
        \begin{center}
        \includegraphics[width=\textwidth]{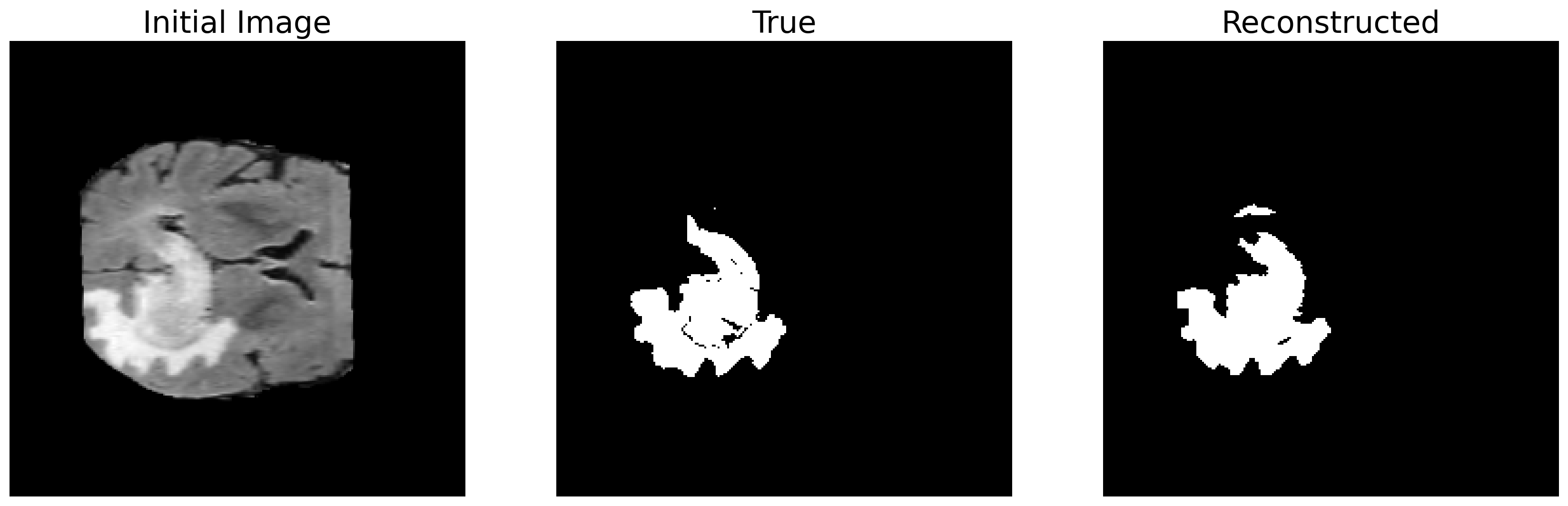}
        \caption{Segmentation results for the “whole tumor” task of the \textit{brain tumor dataset}, obtained with the optimal set of $\Delta_1$, $\Delta_2$ and $\sigma^2$ parameters and with the morphological refinement steps.}
        \label{fig:7}
        \end{center}
        \end{figure}

        In Figure~\ref{fig:7} we present the second segmentation task performed on the \textit{brain tumor dataset} where we apply the method to identify the “whole tumor” region. In this case, the region of interest is characterized by an irregular shape that also contains small holes, concave and convex areas. From the right panel of Figure~\ref{fig:7} we can observe that the reconstructed segmentation mask is very close to the reference one, except of some holes are not entirely identified as well as the small region in the upper area which is misclassified as a tumor part. However, the overall agreement of the two segmentations is good, also considering that the initial image is more complex than the previous ones. The evaluation metric $DSC_{metric}$ reached by the model on the dataset is equal to $0.91$.
        
\begin{figure}
        \centering 
        \subfigure[Healthy subject] {\label{fig:8a}\includegraphics[width=\textwidth]{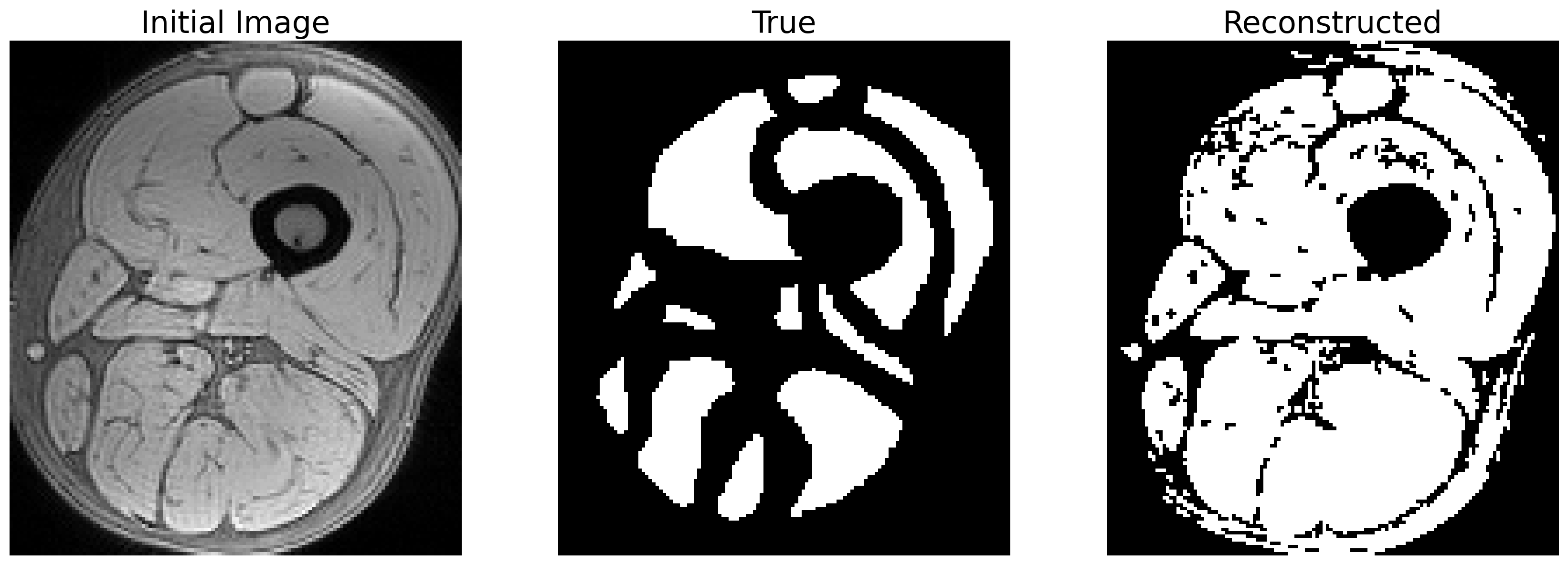}}
        \subfigure[FSHD subject] {\label{fig:8b}\includegraphics[width=\textwidth]{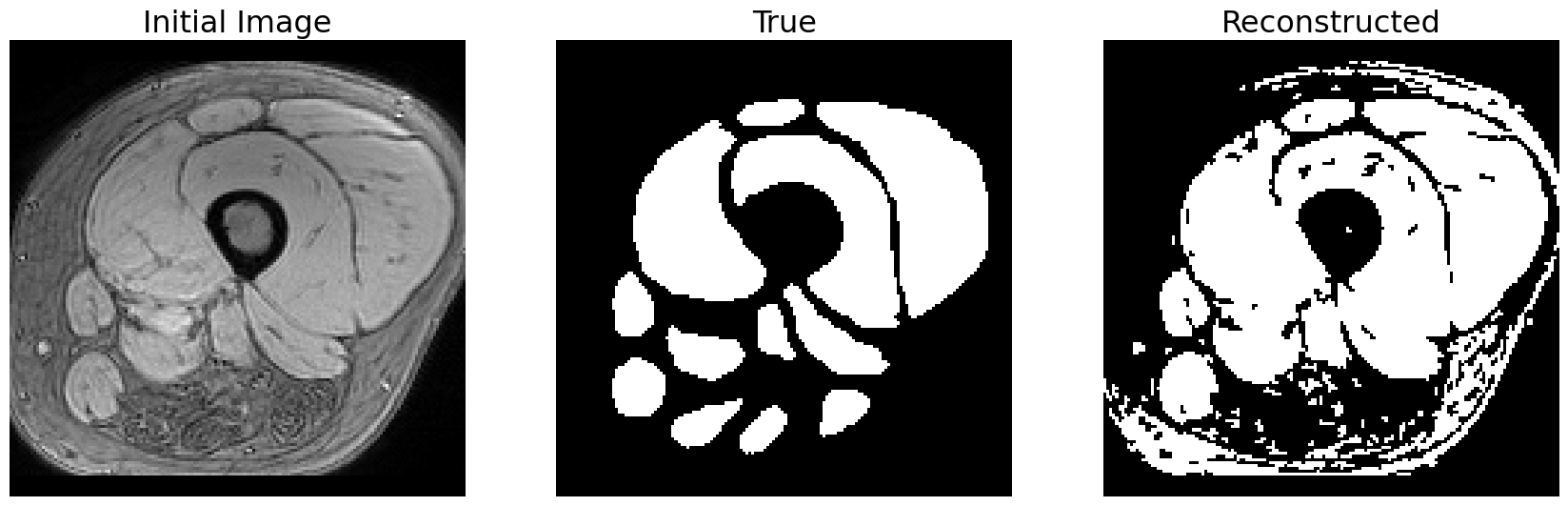}}
        \caption{Results of the segmentation process for the \textit{thigh muscles dataset} obtained with the optimal set of $\Delta_1$, $\Delta_2$ and $\sigma^2$ parameters and with the morphological refinement steps.}
        \label{fig:8}
        \end{figure}

        In Figure~\ref{fig:8} we represent the results of the segmentation process for the \textit{thigh muscles dataset} for a healthy subject~\ref{fig:8a} and a FSHD patient~\ref{fig:8b}. We can observe from the left panel that this dataset, compared to the previous ones, presents some features that appear more complex. Indeed, it is known how the thigh muscles are not well defined structures. Furthermore, thigh muscles often overlap each other or are separated by very thin segments of other tissues. The regions of interest are not uniform in color and differ slightly from the gray level of the surrounding areas. In particular, if we consider the MRI scan of the FSHD subject, we can observe how the disease has altered some muscles (three muscles of the lower area of the image) making them difficult to distinguish from the surrounding adipose tissues. Scan artifacts are stronger than in the other images,  making the homogeneity of the gray levels poorer than in previous ones. 
        
        From the left panel of Figure~\ref{fig:8}, we may observe that the method recognizes the thigh muscles by separating them from the other anatomical structures. However, from a visual assessment we notice that the segmentation precision achieved by the model on this dataset is lower than the previous ones. Especially in the peripheral areas, the method does not correctly distinguish the surrounding tissues from the muscles. If we consider the performances on the FSHD patient image, we can see how the method fails to recognize the muscles altered by the disease, misclassifying them as adipose tissue.
        In terms of the evaluation metric, the model reaches a $DSC_{metric}$ equal to $0.60$ on the healthy subject and a $DSC_{metric}$ equal to $0.73$ for the FSHD one, which are lower values respect the results obtained above.              
        
        It is interesting to observe that, by comparing the results obtained for the healthy and the FSHD patient,  the optimal values of the confidence intervals $\Delta_1$ and $\Delta_2$ are compatible, while the best value of $\sigma^2$ coefficient is greater in the FSHD subject. This happens because the value of the $\sigma^2$ diffusion coefficient is directly related to the inhomogeneities of the gray levels of the pixels, which are greater in the image of the sick patient  due to muscle damage. The increase in the diffusion parameter is particularly interesting as it could be used as an estimate of the FSHD muscle impairment and therefore an indicator of the progress of the disease.
        
        %%==================================%%
        %% PATCH-BASED SEGMENTATION %%
        %%==================================%%
        \subsection{Patch-based 2D biomedical Image segmentation}\label{subsec5.4}
        To alleviate the limitations of the method encountered when dealing with more complex operations and lower quality image data, such as those in the \textit{thigh muscles dataset}, in this section we present an improvement of the segmentation pipeline described in the previous paragraphs.

        This new approach relies on the assumption that the method should recognize fine structures more accurately by focusing on smaller regions of the image. Therefore, we decide to apply the segmentation pipeline~\ref{sec4} and the optimization algorithm~\ref{subsec4.1} to portions of the image that we will call patches (i.e. small subregions of the image defined as two-dimensional pixel arrays) rather than to the whole matrix of pixels~\cite{coupe2011patch, cordier2015patch}.
        
        The method is composed of the following steps. Input images are first converted in square matrices by adding a padding of background pixels on the border of the input array. This step is necessary to obtain square patches, however the segmentation procedure could also be applied to rectangular ones. Images are then divided in non-overlapping patches that are passed as input of the optimization algorithm. In this way, the optimization process will find the best combination of the $\Delta_1$, $\Delta_2$ and $\sigma^2$ parameters for each single patch. Once the local optimization has been performed, the segmentation masks of each patch are estimated through the MC algorithm~\ref{alg:Boltz} with the best combination of parameters. The two refinement steps are applied to each patch masks in order to improve the quality of the results. Hence, all the patch masks are connected together to create the entire segmentation mask and the two post-processing routines are repeated over the complete mask to get the final segmentation.
        
        We test the patch-based segmentation pipeline on the \textit{thigh muscles dataset} for the healthy subject, which was the worst performing example. Using the padding process, we transform the $210 \times 178$ pixel array into a $216 \times 216$ pixel array to fit four square patches to the width and height of the image. In this way, we generate $16$ patches each of $54 \times 54$ pixels.
        Then, for each patch we solve the optimization problem \eqref{eq:min} to determine the optimal parameters $\Delta_1$, $\Delta_2$ and $\sigma^2$ under the constraints $\Delta x\le \Delta_1\le 1.6$, $0.05\le\Delta_2\le 0.2$. The values of $\sigma^2$ are sampled from a log-uniform distribution with support $[e^{-12},1]$ . 
        
\begin{figure}
        \begin{center}
        \includegraphics[width=\textwidth]{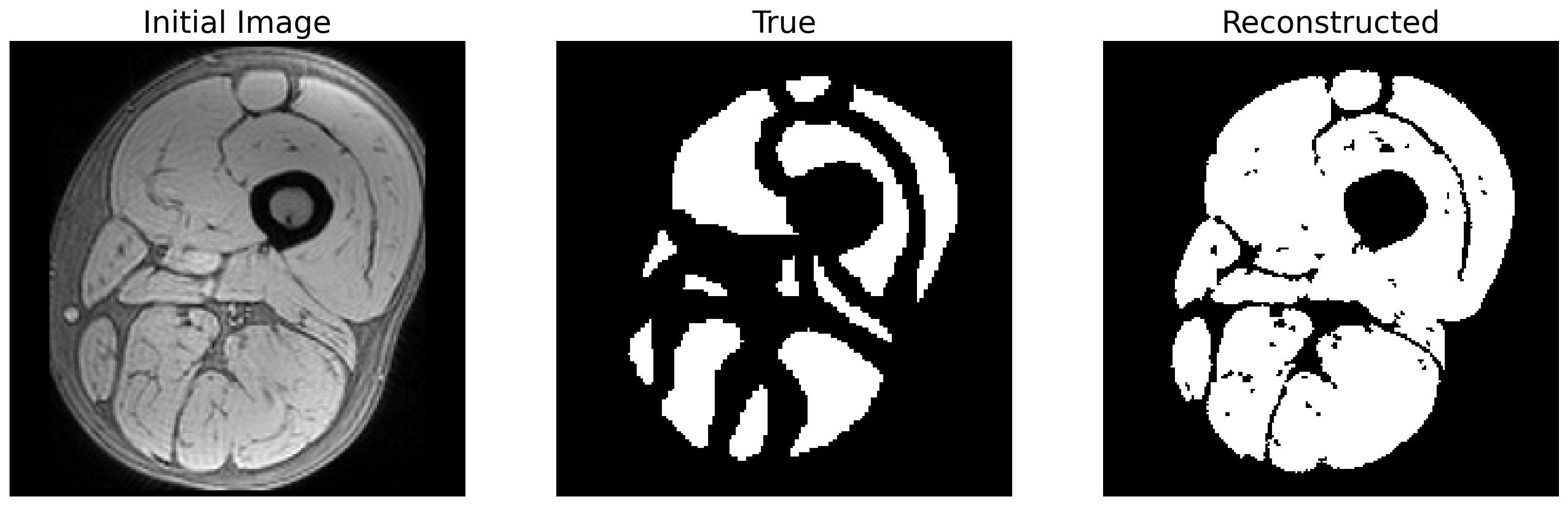}
        \caption{Results of the segmentation process for the \textit{thigh muscles dataset} obtained with the patch-based method with the optimal set of $\Delta_1$, $\Delta_2$ and $\sigma^2$ parameters and the morphological refinement steps.}
        \label{fig:9}
        \end{center}
        \end{figure}

        The results of the patch-based segmentation algorithm for the \textit{thigh muscles dataset} are represented in Figure~\ref{fig:9}. We can observe that the reconstructed mask is clearly closer to the expected one, compared to the mask in Figure~\ref{fig:8}. The method recognizes the thigh muscles with more precision and distinguishes them from the surrounding tissue more accurately. Particularly, if we consider the tissues in the peripheral area of the thigh that were wrongly classified as muscle tissues in Figure~\ref{fig:8}, they are now correctly excluded from the muscles region. In terms of the evaluation metric, the segmentation system achieves a $DSC_{metric}$ of $0.67$. The results obtained from the patch-based method are
appreciable since we successfully excluded subcutaneous adipose tissue from the segmentation mask, which was previously misclassified as part of the muscle. Therefore, the adopted method is capable to improve the separation between the muscle tissue from the surrounding tissue.  
        
        %%==================================%%
        %% DIFFERENT DIFFUSION FUNCTIONS %%
        %%==================================%%
        \subsection{Comparison of different diffusion functions}\label{subsec5.3}
\begin{figure}
        \begin{center}
        \includegraphics[scale = 0.3]{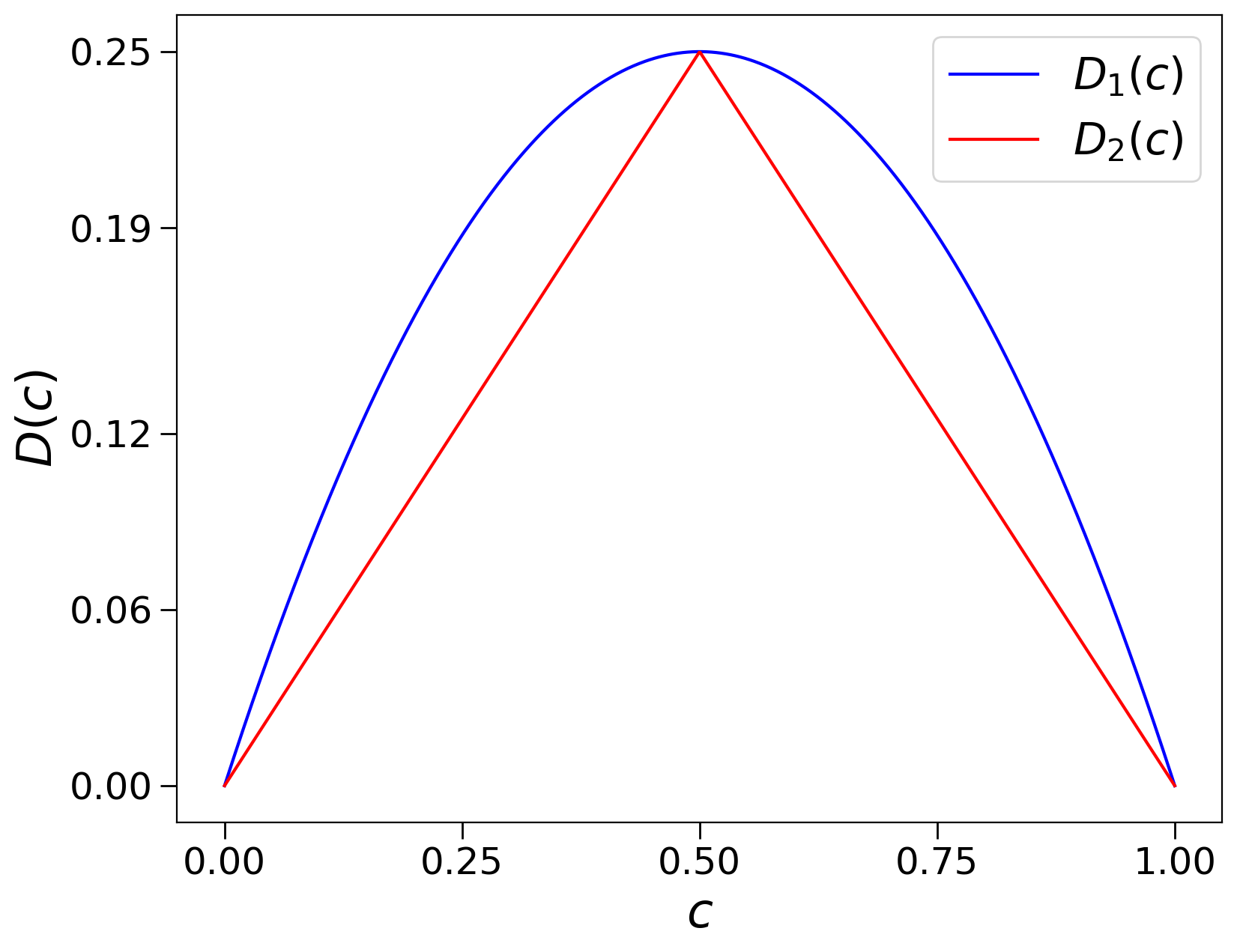}
        \caption{Diffusion functions considered for image segmentation.}
        \label{fig:10}
        \end{center}
        \end{figure}

        In this section we compare the segmentation results obtained with two different local diffusion functions $D(c)$. In details, we consider the functions $D_1(c)>0$ and $D_2(c)>0$ defined as follows
        \begin{equation}
        \label{eq:diff}
        D_1(c) = c(1-c) , \qquad 
        D_2(c) = \begin{cases} \alpha  c &  c\leq0.5\\
        -\alpha  c + \alpha & c>0.5
        \end{cases}
        \end{equation}
        with $c \in [0,1]$ and $\alpha=\frac{1}{2}$ such that $D_1\left(\frac{1}{2}\right)=D_2\left(\frac{1}{2}\right)$, see Figure~\ref{fig:10}. 
        
       Hence, we implemented the DSMC Algorithm~\ref{alg:Boltz} to segment the tumor core region of the \textit{brain tumor dataset} using both $D_1(c)$ and $D_2(c)$. The parameters $\Delta_1$, $\Delta_2$ and $\sigma^2$ were optimized independently for both cases, and the best configurations are reported in Table~\ref{table:best2}. {We may observe that, consistently with the fact that for all $c\in [0,1]$ we have $D_1(c)\ge D_2(c)$, the estimated multiplicative diffusion coefficient is larger for $D_2(c)$.}
        
        \begin{table}
        \centering
        \begin{tabular}{  c  c c  c } 
        \hline
        $\mathbf{D(c)}$ & $\mathbf{\Delta_1}$ & $\mathbf{\Delta_2}$ & $\mathbf{\sigma^2}$ \\ 
        \hline 
        $D_1$ & 0.69 & 0.17 & 0.03 \\
        $D_2$ & 0.58 & 0.27 & 0.06 \\
        \hline
        \end{tabular}
        \caption{Estimated values of $\Delta_1$, $\Delta_2$ and $\sigma^2$ by the optimization process for two different diffusion functions.}
                \label{table:best2}

        \end{table}

\begin{figure}
        \begin{center}
        \includegraphics[width=\textwidth]{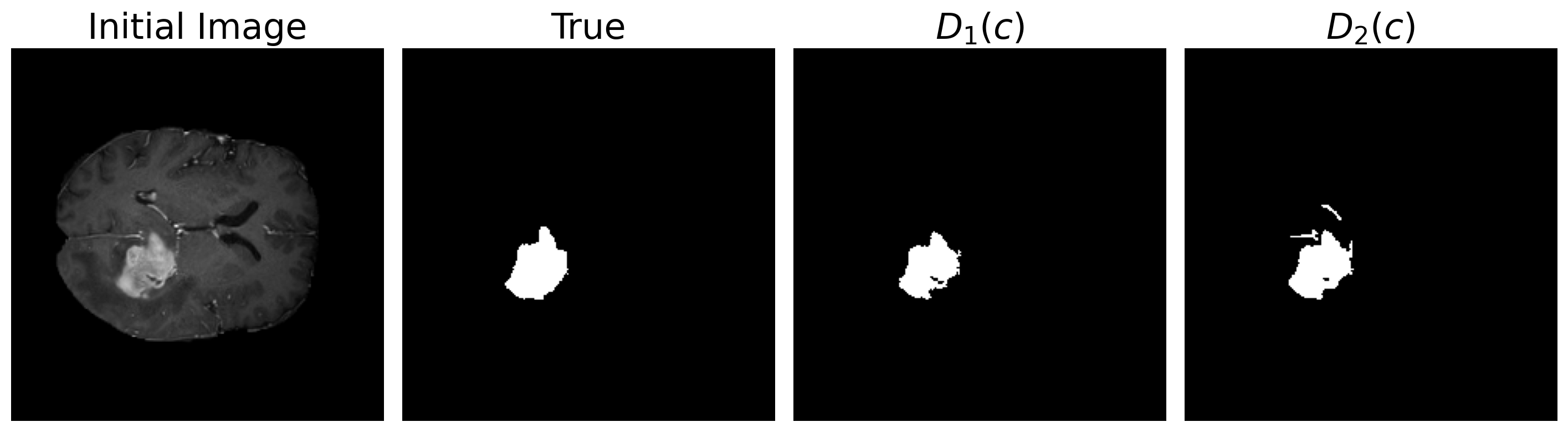}
        \caption{Segmentation results for the “tumor core” task of the \textit{brain tumor dataset}, obtained with the optimal set of $\Delta_1$, $\Delta_2$ and $\sigma^2$ parameters and with the two different diffusion functions $D_1$ and $D_2$.}
        \label{fig:11}
        \end{center}
        \end{figure}

       In Table~\ref{table:dsc2} we report the segmentation results in terms of $DSC_{metric}$ obtained with local diffusion functions $D_1(c)$ and $D_2(c)$. We may observe that we obtained better performance of the segmentation pipeline defined with $D_1(c)$ with respect to the case $D_2(c)$.  It is important to observe also that the difference in performance are alleviated by the morphological refinement step. Figure~\ref{fig:11} shows the segmentation results obtained with both diffusion functions. We may observe how the segmentation mask obtained with $D_2$ includes two small regions erroneously classified as part of the tumor area. These results suggest that the segmentation performance is influenced by the selection of a different diffusion function and, in this specific example, the quadratic function $D_1(c)$ leads to better results in the considered case.

    {However, the choice of the optimal diffusion function may be application-specific and be related to various factors like the nature of the images, their sizes, compactness and homogeneity of the edges of the segmentation region. All these factors could influence the choice of the optimal diffusion function for the specific application.}

        \begin{table}[h]
        \centering
        \begin{tabular}{ l c c c} 
        \hline 
        & \multicolumn{1}{l}{$\mathbf{DSC_{metric}}$} & \multicolumn{1}{l}{$\mathbf{DSC_{metric}}$}\\
        \textbf{Diffusion function} & \textbf{Row mask} & \textbf{Final mask} \\ 
        \hline 
        $D_1$ & 0.87 & 0.93 \\
        $D_2$ & 0.79 & 0.90 \\
        \hline
        \end{tabular}
                \caption{Values of $DSC_{metric}$ before (row mask) and after (final mask) the morphological refinement steps.}
        \label{table:dsc2}

        \end{table}

        %%==================================%%
        %% CONCLUSIONS %%
        %%==================================%%
        \section*{Conclusions}\label{sec6}
        
        In this work we proposed a kinetic model for consensus-based segmentation tasks inspired by the Hegselmann-Krause (HK) model.  We derived a Boltzmann description of the generalized HK model based on a binary interaction scheme for the particles' locations and features. In the quasi-invariant limit we showed the consistency of the approach with existing mean-field modelling approaches by deriving a surrogate Fokker-Planck model with nonlocal drift. The Boltzmann-type description allows us to apply efficient direct simulation Monte Carlo schemes to evaluate the collision dynamics with low computational burden. Hence, we proposed an optimization strategy for the internal parameter configuration.  {The optimization task is completed by considering a $DSC_{loss}$ in view of its computational  efficiency. }
        
        This model-based segmentation strategy is tested on three different biomedical datasets: a HL60 cell nuclei dataset, a brain tumor dataset and a thigh muscles dataset. The performances of the method are evaluated in terms of the $DSC_{metric}$ that quantifies the overlap between the reconstructed mask and the reference mask. Good results are obtained for the HL60 cell nuclei dataset and the brain tumor dataset, while for the thigh muscle dataset the segmentation accuracy is lower. For this more complex dataset, we proposed a patch-based approach which consists of dividing the image into smaller arrays of pixels and applying the segmentation system to these subregions of the initial image. This second version of the method improves the quality of the segmentation mask. 

Although satisfactory results have been achieved, this work can be improved in different directions. Firstly, our work primarily focuses on validating the applicability of the segmentation model to real biomedical images. To enhance the robustness and generalizability of our findings, increasing the sample size of our study and testing the proposed methodology on other imaging modalities and anatomical regions would be relevant. Additionally, exploring the potential of generating data-driven segmentations would further enhance the relevance of our method, enabling its application in scenarios where ground truth masks are not readily available. Furthermore, the aim of this study is to develop a versatile segmentation method applicable to a broad spectrum of biomedical images and segmentation tasks, rather than being tailored to a specific imaging question. The segmentation method can be further refined to address the specific needs of diverse tasks, potentially enhancing its flexibility and effectiveness.
        
        Several extensions of the presented modelling approach to include color images and modified interaction functions are currently under study and will be discussed in future works. {Additionally, we plan to systematically compare various metrics for optimizing the model parameters, and we intend to develop a pipeline for more accurately selecting the diffusion weight.}

    \section*{Acknowledgements}
    This work has been written within the activities of the GNFM group of INdAM (National Institute of High Mathematics). M.Z. acknowledges the support of Next Generation EU Project No.P2022Z7ZAJ (A unitary mathematical framework for modelling muscular dystrophies).
    R.F.C., S.F. and A.L thank the INFN-CSN5 research project \textit{next$\_$AIM (Artificial Intelligence in Medicine: next steps)}, \url{https://www.pi.infn.it/aim/}. A.L. thanks the national project ``MIUR, Dipartimenti di Eccellenza Program (2018-2022), project F11I18000680001".
    
    \bibliography{refs}
	
\end{document}